\documentclass[apj]{emulateapj}

\usepackage{xspace,graphics,graphicx}
\usepackage{lscape,natbib}
\slugcomment{Submitted to ApJ} 

\newcommand{\thresh}{$\rm \Sigma_{SFR} \gtrsim 0.1~M_{\odot}~yr^{-1}~kpc^{-2}$}
\newcommand{\thresheq}{$\rm \Sigma_{SFR} = 0.1~M_{\odot}~yr^{-1}~kpc^{-2}$}
\newcommand{\mkms}{{\rm \; km\;s^{-1}}}
\newcommand{\kms}{km~s$^{-1}$ }
\newcommand{\kmsns}{km~s$^{-1}$}
\newcommand{\prewavesize}{529 }
\newcommand{\fullsampsize}{468 }
\newcommand{\bcsampsize}{407 }
\newcommand{\gdlir}{213 }

\shorttitle{Galactic Winds at $0.7 < z < 1.5$}
\shortauthors{Rubin et al.}

\begin{document}

\title{The Persistence of Cool Galactic Winds in High Stellar Mass Galaxies Between $z \sim 1.4$ and $\sim 1$\altaffilmark{1}}
\author{Kate H. R. Rubin\altaffilmark{2,3}, Benjamin J. Weiner\altaffilmark{4}, David C. Koo\altaffilmark{2}, Crystal L. Martin\altaffilmark{5,6}, 
J. Xavier Prochaska\altaffilmark{2}, Alison L. Coil\altaffilmark{7} \& Jeffrey A. Newman\altaffilmark{8}}







\altaffiltext{1}{Some of the data presented herein were obtained at the W. M. Keck Observatory, which is operated as a scientific partnership among the California Institute of Technology, the University of California and the National Aeronautics and Space Administration.  The Observatory was made possible by the generous financial support of the W.M. Keck Foundation.}
\altaffiltext{2}{University of California Observatories, University of California, Santa Cruz, CA 95064} 
\altaffiltext{3}{rubin@ucolick.org}
\altaffiltext{4}{Steward Observatory, 933 N. Cherry St., University of Arizona, Tucson, AZ 85721}
\altaffiltext{5}{Department of Physics, University of California, Santa Barbara, CA, 93106}
\altaffiltext{6}{Packard Fellow}
\altaffiltext{7}{Department of Physics, University of California, San Diego, CA, 92093}
\altaffiltext{8}{Department of Physics and Astronomy, University of Pittsburgh, Pittsburgh, PA 15260}

\begin{abstract}
We present an analysis of the \ion{Mg}{2} $\lambda \lambda 2796,2803$
and \ion{Fe}{2} $\lambda \lambda 2586, 2600$ absorption line profiles in 
coadded spectra of \fullsampsize galaxies at $0.7 < z < 1.5$.  
The galaxy sample, drawn from the Team Keck Treasury Redshift 
Survey of the GOODS-N field, has a range in stellar mass ($\rm M_*$) comparable to that of the sample at $z \sim 1.4$ analyzed in a similar manner by \citet[][W09]{Weiner2009}, but extends to lower redshifts and has specific star formation rates which are lower by $\sim 0.6$ dex.  We identify outflows of cool gas from the Doppler shift of the \ion{Mg}{2} absorption lines and find that the equivalent width (EW) of absorption due to outflowing gas increases on average with $\rm M_*$ and star formation rate (SFR).  We attribute the large EWs measured in spectra of the more massive, higher-SFR galaxies to optically thick absorbing clouds having large velocity widths.  The outflows have hydrogen column densities $N(\mathrm{H}) \gtrsim 10^{19.3} \rm cm^{-2}$, and extend to velocities of $\sim 500 \mkms$.  While
galaxies with $\rm SFR > 10~M_{\odot}~yr^{-1}$ host strong outflows in both this and the W09 sample, we do not detect outflows in lower-SFR (i.e., $\rm \log M_*/M_{\odot} \lesssim 10.5$) galaxies at lower redshifts.  
Using a simple galaxy evolution model which assumes exponentially declining SFRs, we infer that strong outflows persist in galaxies with $\rm \log M_*/M_{\odot} > 10.5$ as they age between $z = 1.4$ and $z \sim 1$, presumably because of their high absolute SFRs.  
Finally, using high resolution HST/ACS imaging in tandem with our spectral analysis, we find evidence for a weak trend (at 1$\sigma$ significance) of increasing outflow absorption strength with increasing galaxy SFR surface density.

\end{abstract}

\keywords{galaxies: absorption lines --- galaxies: evolution --- galaxies: ISM}

\section{Introduction}\label{sec.intro}

Galactic-scale gaseous outflows are a basic element of the process of galaxy evolution, and are observed in galaxies in the local Universe and out to $z \sim 6$ \citep[e.g.,][]{Heckman2000,Frye2002,Ajiki2002,Shapley2003,Martin2005,Rupke2005b,Veilleux2005,Weiner2009}.  
They are invoked to explain a wide variety of observational and theoretical results concerning the evolving stellar and gaseous content of dark matter halos.  
Outflows are a key component of the theory of merger-driven galaxy evolution, in which the primary mechanism driving the observed increase in the number density of ``red and dead" galaxies from $z \sim 1$ to today is the merging of gas-rich blue galaxies \citep[e.g.,][]{Faber2007,Tremonti2007,Hopkins2008,Sato2009}.  The removal (i.e., outflow) of gas is an expected consequence of merging, and is a favored means by which star formation is subsequently quenched in the remnant.  
Outflows must be incorporated into models of disk galaxy formation which correctly reproduce the observed scaling relations between disk rotation velocity, stellar mass ($\rm M_*$), and radius \citep[e.g.,][]{Dutton2009}.
Finally, galactic winds may give rise to the \ion{Mg}{2}-absorbing gas in the extended gaseous halos around galaxies observed along QSO sightlines.  Several lines of evidence support this idea \citep[e.g.,][]{Bond2001,Bouche2006,Bouche2007,MenardChelouche2009}; however, alternative origins for this gas have also been suggested, i.e., multiphase cooling of hot halo gas, accretion of cold gas from the intergalactic medium, or tidal stripping \citep{MB2004,TC2008,Wang1993}.  

In spite of their importance, 
the physical processes responsible for driving galactic outflows are not well understood.   Energy from supernova explosions or AGN feedback heats the surrounding gas and may displace both hot and cool gas originating in the interstellar medium (ISM), possibly removing it to the galactic halo or beyond.  Momentum deposition from radiation or cosmic ray pressure may also contribute to the acceleration of cool ISM gas; the relative importance of these two processes is hotly debated \citep{DekelSilk1986,Murray2005,Socrates2008,StricklandHeckman2009}.  In galaxies which are known to host outflows in the local Universe, the hot phase is observed in X-ray emission, while the cooler phase is detected via optical emission lines (e.g., H$\alpha$, [\ion{N}{2}] $\lambda \lambda 6548, 6583$) and in absorption against the stellar continuum.  However, it remains difficult to predict whether galactic winds will
form and whether they will affect the kinematics of the ISM in a given galaxy.   
Theoretical studies propose that there is a ``threshold" star formation rate surface density ($\rm \Sigma_{SFR}$) below which supernova-driven superbubbles cannot break out of a galactic disk and form a wind \citep[e.g.,][]{McKeeOstriker1977}.  
Observational constraints on the value of a universal threshold $\rm \Sigma_{SFR}$ 
remain merely suggestive \citep[e.g.,][]{LehnertHeckman1996,Martin1999,Heckman2002,Dahlem2006}.

For instance, several studies have identified outflows in galaxies in which absorption line transitions tracing cool gas are blueshifted with respect to the systemic velocity.
The \ion{Na}{1} D $\lambda \lambda$ 5890, 5896 doublet traces the kinematics and column of gas at $\rm T \sim 100-1000~K$, revealing outflows in local dwarf starbursts \citep{SchwartzMartin2004} and luminous infrared galaxies (LIRGs) out to $z \sim 0.5$ \citep{Heckman2000,Martin2005,Rupke2005b}.  
UV spectroscopy of both low and high-ionization transitions such as \ion{Si}{2} $\lambda 1260$ and \ion{C}{4} $\lambda \lambda 1548, 1550$ in Lyman Break Galaxies (LBGs) at $z \sim 3$ has uncovered outflows of hundreds of \kms \citep[e.g.,][]{Shapley2003} in these objects.
All of these galaxies have high spatial concentrations of star formation, with $\rm \Sigma_{SFR} > 0.1 \rm ~M_{\odot}~yr^{-1}~kpc^{-2}$ \citep{Heckman2002}; the nearby galaxies have $\rm \Sigma_{SFR}$ values among the largest in the local Universe.  
If there is a ``threshold" $\rm \Sigma_{SFR}$ required to drive outflows, this suggests that it is likely equal to or below $0.1 \rm ~M_{\odot}~yr^{-1}~kpc^{-2}$ for local galaxies.  The value of a threshold $\rm \Sigma_{SFR}$ for galaxies in the distant Universe remains poorly constrained, although the high $\rm \Sigma_{SFR}$ ($> 1 \rm ~M_{\odot}~yr^{-1}~kpc^{-2}$) values in LBGs \citep{Steidel1996,Meurer1997} suggest a similarly low threshold at $z \sim 3$.

Not only are outflows expected to contribute to increased numbers of post-merger, ``quenched" red galaxies, 
but they may also influence the gradual decline of star formation since $z \sim 1$ among blue galaxies \citep{Noeske2007a,Sato2009}, as this decline is likely driven by gas exhaustion.
The presence of winds in both distant star-forming and red sequence galaxies has been established \citep[e.g.,][hereafter W09]{Sato2009,Weiner2009}, although the co-evolution of outflows with the buildup of $\rm M_*$ and the decreasing cosmic star formation rate (SFR) density \citep[e.g.,][]{Hopkins2004} remains unexplored at $z \gtrsim 0.5$.

We use a sample of \fullsampsize galaxies at $0.7 < z < 1.5$ drawn from the Team Keck Treasury Redshift Survey \citep[TKRS;][]{Wirth2004} of the GOODS-N field to examine the kinematics of cool ($\rm T \lesssim 10^4~K$) gas traced by \ion{Mg}{2} and \ion{Fe}{2} absorption in coadded spectra.  Because LIRG-like levels of star formation occur in most massive star-forming galaxies at $z \sim 1$ \citep{Noeske2007a}, and because cool outflows are common among LIRGs in the local Universe, the galaxies in our sample are prime candidates for hosting outflows \citep{Rupke2005b,Heckman2000}.  Indeed, W09 showed via analysis of coadded spectra of $\sim 1400$ blue, star-forming galaxies at $z \sim 1.4$ that outflows traced by \ion{Mg}{2} absorption are very common among the objects in their sample, and further demonstrated that outflow velocities and absorption strengths increase with both $\rm M_*$ and SFR over the ranges $\rm 9.5 < \log M_* / M_{\odot} < 10.8$ and $\rm 10~ M_{\odot}~yr^{-1} < SFR < 50~M_{\odot}~yr^{-1}$.

However, additional evidence indicates that the outflows observed in galaxies at $z \sim 1.4$ by W09 must cease by $z \sim 0.5$.  W09 suggest that many of the galaxies in their sample will evolve into massive spirals (rather than ellipticals) at $z \sim 0$ \citep{Blanton2006,Noeske2007b}.  
The study of \ion{Na}{1} kinematics in both star-forming and quiescent, red-sequence galaxies at $0.11 \le z \le 0.54$ by \citet{Sato2009} 
showed that outflows with velocities $> 50~\rm km~s^{-1}$ traced by \ion{Na}{1} are detected in $\sim 80\%$ of the blue cloud galaxies in their sample, while the remaining 20\% of these galaxies lack outflows.  Although the \citet{Sato2009} sample of blue cloud galaxies is incomplete, this suggests that at least some star-forming galaxies at $z \sim 1.4$ cease to drive winds in the redshift range $0.5 < z < 1.4$.  

We use our sample to explore the dependence of the equivalent width (EW) of outflowing gas on galaxy $\rm M_*$ and SFR and examine the relative importance of these two properties in determining outflow absorption strength among star-forming objects.  We follow the co-evolution of outflows, galaxy $\rm M_*$ and SFR in the range $0.7 \lesssim z \lesssim 1.4$ and search for the expected decline in outflow EW with increasing galaxy age.
Finally, using coadded spectra in concert with the deep HST/ACS imaging available in the GOODS-N field, we study the relationship between $\rm \Sigma_{SFR}$ and the EW of absorption due to outflowing gas for the first time beyond the local Universe. 

We describe our sample, spectroscopy and imaging data in \S2.  We show evidence for outflow in a few individual galaxy spectra and in a coadded spectrum in \S3, and describe the measurements we use to quantify outflow properties in \S4.  Section 5 describes our derivations of galaxy properties, and \S6 examines trends in outflow properties with galaxy SFR, $\rm M_*$, SFR surface density ($\rm \Sigma_{SFR}$) and redshift.  Section 7 discusses \ion{Fe}{2} absorption properties.  Sections 8 and 9 contain a discussion of these results and our conclusions.  We adopt a $\Lambda$CDM cosmology with $H_0 = 70~\rm km~s^{-1}~Mpc^{-1}$, $\rm \Omega_M = 0.3$, and $\rm \Omega_{\Lambda} = 0.7$.  Magnitudes quoted are in the AB system unless otherwise specified, and stellar masses are reported in units of $\rm M_{\odot}$.

\section{Data, Sample, and Stacking Technique}
\subsection{Spectra}\label{sec.spectra}
We use publicly available spectra from the TKRS \citep{Wirth2004} for analysis of \ion{Mg}{2} and \ion{Fe}{2} kinematics.  In brief, the TKRS is a magnitude-limited spectroscopic survey of galaxies selected to have $R_{AB} < 24.4$ in the GOODS-N field.  Spectra were obtained using one hour exposures with DEIMOS \citep{Faber2003} on the Keck 2 telescope.  The $600~\rm l~mm^{-1}$ grating blazed at $7500\rm ~\AA$ was used with a 1\arcsec \ wide slit.  Spectra have  
a resolution of $\sigma_{inst} = 1.4$ \AA \ \citep{Weiner2006}, and cover $\sim 4600-9800\rm~\AA$ with $\rm 0.648~\AA ~pix^{-1}$.  Approximately 1440 spectra of galaxies and AGN were obtained in total.  

The spectra were reduced using an early version of the pipeline developed by the DEEP2 Redshift Survey Team \citep[][J. Newman et al.\ 2009, in preparation]{Coil2004}.  To determine first pass redshifts, the pipeline calculated the best fitting source spectrum at each lag position, or redshift, from a linear combination of several template eigenspectra (Strauss et al. 2002; Glazebrook et al. 1998).  The template spectra included stars, galaxy absorption and emission line spectra, and an AGN spectrum.  The 10 best fits (i.e., with the smallest $\chi^2$ values) were saved, and members of the TKRS team then visually checked the best solution, replacing it in some cases, and assigned it a quality code.
\citet{Wirth2004} performed a detailed comparison between TKRS redshift determinations and redshifts measured in other surveys, and estimated that the uncertainty in TKRS redshifts is $\sim 100~\rm km~s^{-1}$ or better.

We limit our sample to spectra with high redshift quality codes, i.e., with confidence level $> 90\%$, and with coverage of the 
\ion{Mg}{2} $\lambda \lambda 2796,2803$ doublet.  
We then visually inspect each of these spectra to check for broad \ion{Mg}{2} emission lines; we exclude four objects with broad emission from our analysis (with object IDs 5609, 3660, 1488 and 9377).  These cuts limit the sample size to 625 objects.  

As noted in \citet{CowieBarger2008}, some of the TKRS spectra suffer from poor sky subtraction, and spectra of faint objects can have negative continua.  This implies that there are large systematic errors not accounted for in the one-dimensional error array for each spectrum.  If the distribution of the errors in the estimates of the sky level were symmetric, the sky subtraction errors would not affect our results, as we work primarily with coadded spectra.  However, negative continua (i.e., oversubtractions) occur preferentially when there is excess scattered light in a given slit, whereas we do not expect that undersubtraction results from a specific systematic issue.  Thus, we expect that overestimation of the sky level is more likely to occur than underestimation in general.
Including the objects with oversubtracted sky in coadded spectra will tend to reduce the continuum level of the coadd.  We exclude the 96 spectra with a negative median continuum level measured between rest wavelengths 2790 \AA \ and 2810 \AA \ (i.e., 15\% of the sample).  This method of tagging spectra with poor sky subtraction is almost certainly not comprehensive; however, it likely removes the worst cases from the sample.

\begin{figure}
\includegraphics[width=3.25in,angle=90]{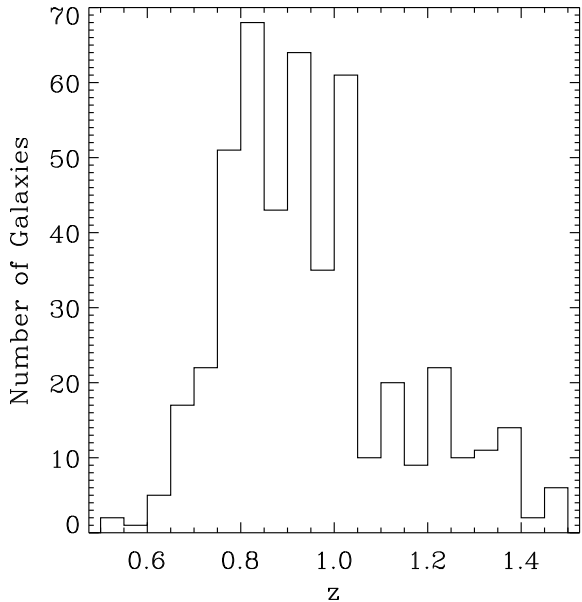}
\caption{Histogram showing redshift distribution of all \fullsampsize galaxies included in our analysis in bins of $\Delta z = 0.05$.  Only spectra with coverage of \ion{Mg}{2} $\lambda \lambda 2796,2803$ are included, which places a lower bound on the redshift distribution at $z \approx 0.7$. \label{fig.redshiftdist}}
\end{figure}

Among the remaining \prewavesize spectra, a few have very poor wavelength solutions in the blue due to the paucity of arc lamp lines at $\lambda < 5000$ \AA.  This may affect the accuracy of the redshift determination as well as the offset of UV absorption lines from systemic velocity.  In order to assess the quality of the wavelength solutions, we fit a single Gaussian to the sky lines at $5197~\rm\AA$ and $5200~\rm\AA$, which are not resolved in the TKRS spectra.  We define $\rm v = 0~km~s^{-1}$ to be at the average of the true wavelengths of the two lines ($\langle \lambda_{5197,5200} \rangle$) and measure the velocity offset of the sky line complex in each spectrum.  
Our assumption that the centroid of a single Gaussian fit to the blended lines will be at the average of their central wavelengths holds if the lines have equal strengths.  In the extreme case in which one of the lines is completely absent, an additional offset of $\pm 68\rm~km~s^{-1}$ must be added to the calculated offset.  To find the mean velocity offset of the sky line complex for the sample ($\langle v_{obs,5197+5200}\rangle$), we first exclude spectra with sky lines offset by more than $200~\rm km~s^{-1}$ from $\langle\lambda_{5197,5200} \rangle$, and then calculate the mean of the velocity offset in the remaining spectra.  The resulting $\langle v_{obs,5197+5200} \rangle$ is $13~\rm km~s^{-1}$.  We then exclude all spectra from the sample (61 objects) with sky lines offset by more than $130 \rm~km~s^{-1}$ from $\langle v_{obs,5197+5200} \rangle$.  Accounting for variations in the relative strengths of the sky lines, this means that all remaining \fullsampsize spectra have sky lines offset by no more than $211\rm~km~s^{-1}$ from their true central wavelength; these spectra make up our final sample.   
Figure~\ref{fig.redshiftdist} shows the redshift distribution of the objects in the sample.  The median redshift of the total sample is $z = 0.94$.

The standard deviation of the distribution in sky line velocity offsets for spectra remaining in the sample is $40 \rm ~km~s^{-1}$ (not accounting for variations in the strengths of the sky lines).  This dispersion measures the typical wavelength uncertainty in the TKRS spectra at $\lambda \sim 5200$ \AA.  We estimate considerably smaller uncertainties at redder wavelengths, as arc lines are more abundant in this range.  For instance, a Gaussian fit to the sky line at 5577 \AA \ yields maximum velocity offsets $< 30~\rm km~s^{-1}$ and a dispersion of $10~\rm km~s^{-1}$ for the spectra in the sample.  

The errors in the wavelength solution introduces uncertainties in the systemic velocity of the UV absorption lines in each spectrum.  Throughout the paper, we assume that the systemic velocity of each galaxy is given by the TKRS redshift.  As redshift determinations are most strongly dependent on sections of the spectra redward of observed wavelength $\rm \lambda_{obs} \sim 5577$ \AA \ in the redshift range of our sample, they should not be affected by poor blue wavelength solutions.  However, the UV absorption lines of interest are blueward of $\rm \lambda_{obs} \sim 5577$ \AA \ in the redshift range 
$z < 1$ in the case of \ion{Mg}{2} and in the range $z < 1.15$ in the case of \ion{Fe}{2}.  We estimate the error in the systemic velocity for these transitions by assuming the wavelength solution is perfect at $\rm \lambda_{obs} \sim 5577$ \AA \ and that it varies linearly to the blue, with absolute velocity offsets at $\rm \lambda_{obs} \sim 5200$ \AA \ between $\rm 20~km~s^{-1}$ and $\rm 200~km~s^{-1}$.  
At the very bluest wavelengths, the velocity offset is $< 135~\rm km~s^{-1}$ for sky line velocity offsets $< 40~\rm km~s^{-1}$, and have a maximum of $\rm 675~km~s^{-1}$ for sky line velocity offsets of $200 \mkms$.  Near the median redshift of the sample ($z = 0.94$), the velocity offset at \ion{Mg}{2} is $\sim 27~ \mkms$ for sky line velocity offsets of $40~ \mkms$.
These errors in the systemic velocities of UV absorption lines must be considered when interpreting results from coadded spectra, as discussed in \S\ref{sec.stacking}.

\subsection{Imaging}
We use the high quality HST/ACS imaging available in the GOODS-N field \citep{Giavalisco2004}.  The imaging covers a $10\arcmin \times 16\arcmin$ area with the ACS F435W, F606W, F814W and F850LP bands ($B_{435}$, $V_{606}$, $i_{775}$ and $z_{850}$).  The limiting surface brightness at $1\sigma$ in a $\rm 1~sq. \arcsec$ aperture in the F850LP band is $\mu_{AB} = 27.3~\rm mag~arcsec^{-2}$ \citep[][version 1 release]{Giavalisco2004}.  We use the mosaic data in each band with a pixel scale $\rm 0.03\arcsec ~ pix^{-1}$.

\subsection{Stacking Technique}\label{sec.stacking}
We use code written by J.A.N. to coadd TKRS spectra. 
The code first masks out bad pixels in each object spectrum.  
It then renormalizes each inverse variance array so that it has a median equal to 1.  Each spectrum and its associated inverse variance array is linearly interpolated onto a grid of rest-frame wavelength, and the spectra are coadded.  The flux in each pixel is weighted by the renormalized inverse variance, so that pixels with more noise from sky emission are given less weight, while  
each spectrum overall contributes to the stack in proportion to its flux and it does not have an extra weighting corresponding to its $\rm S/N$.  This results in a coadd which is ``light-weighted".  This method of coaddition is the same as that used in W09.

Errors in TKRS redshifts as well as errors in the wavelength solution will have the effect of broadening the absorption and emission features located at the true systemic velocity in the coadded spectra.  The redshift uncertainties discussed above are $< 100~\rm km~s^{-1}$, while velocity offsets due to poor wavelength solutions are estimated to be $< 135~\rm km~s^{-1}$ for the 68\% of the spectra with the smallest sky line velocity offsets, making the extreme assumption that the UV absorption lines of interest are always at the bluest end of the wavelength coverage.  
We therefore expect that absorption observed at velocities offset from systemic by $\gtrsim 200~\rm km~s^{-1}$ in both the \ion{Mg}{2} and \ion{Fe}{2} transitions in the coadd is most likely due to absorption at velocities offset from the true systemic velocity of the galaxies in the sample.

\section{Rest-frame UV Metal-line Absorption in TKRS Galaxies}
\subsection{Detection of Outflowing Gas in TKRS2158}\label{sec.TKRS2158}
The continuum $\rm S/N$ of a typical TKRS spectrum near \ion{Mg}{2} is $\sim 1~\rm pixel^{-1}$.  We are therefore not able to visually identify \ion{Mg}{2} absorption in the vast majority of the individual spectra.  There are a few brighter galaxies, however, for which rest-frame UV absorption lines are evident.     

\begin{figure*}
\includegraphics[width=2.4in]{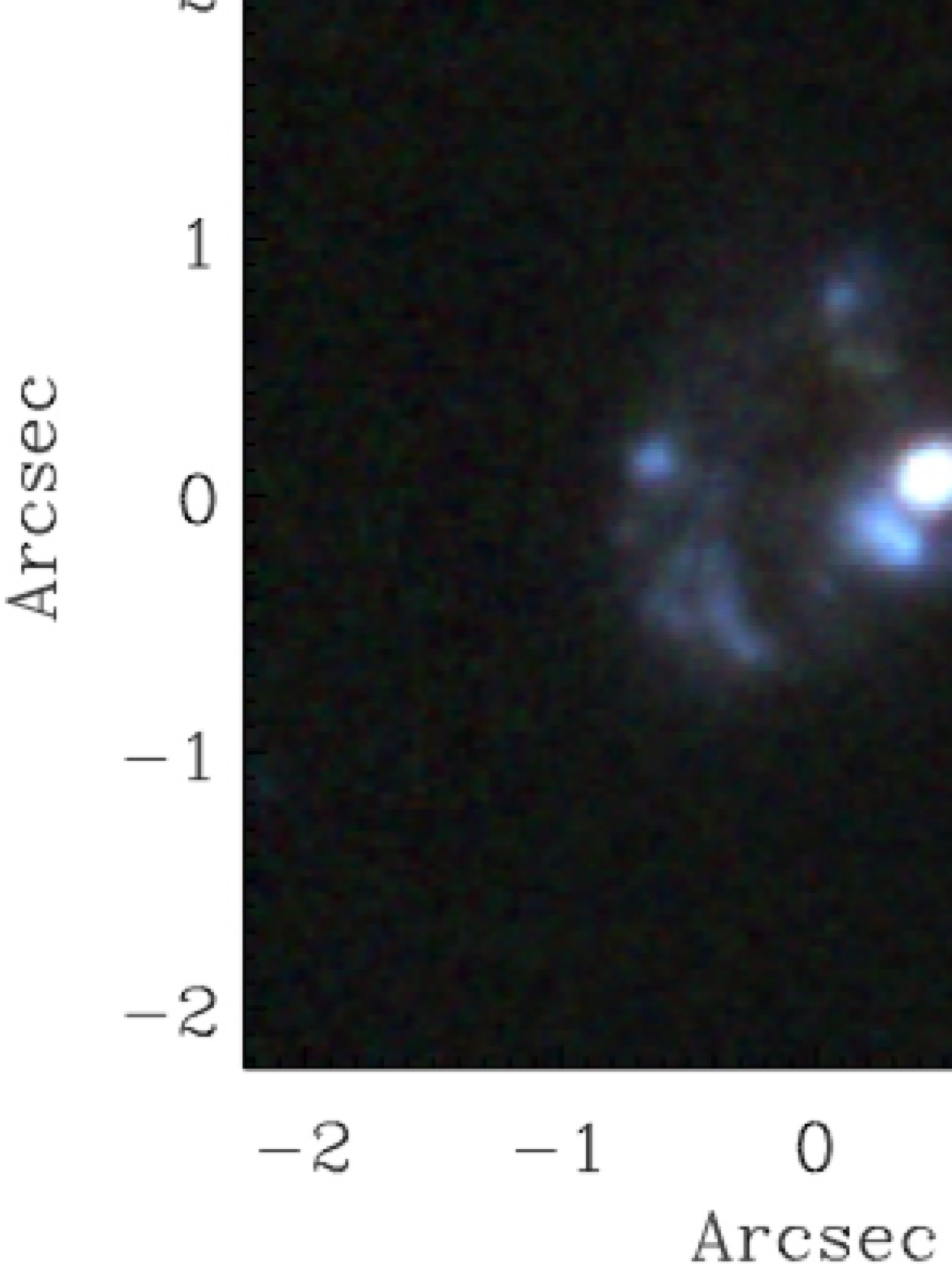}
\includegraphics[angle=90,width=4.75in,viewport=-5 -10 380 750,clip]{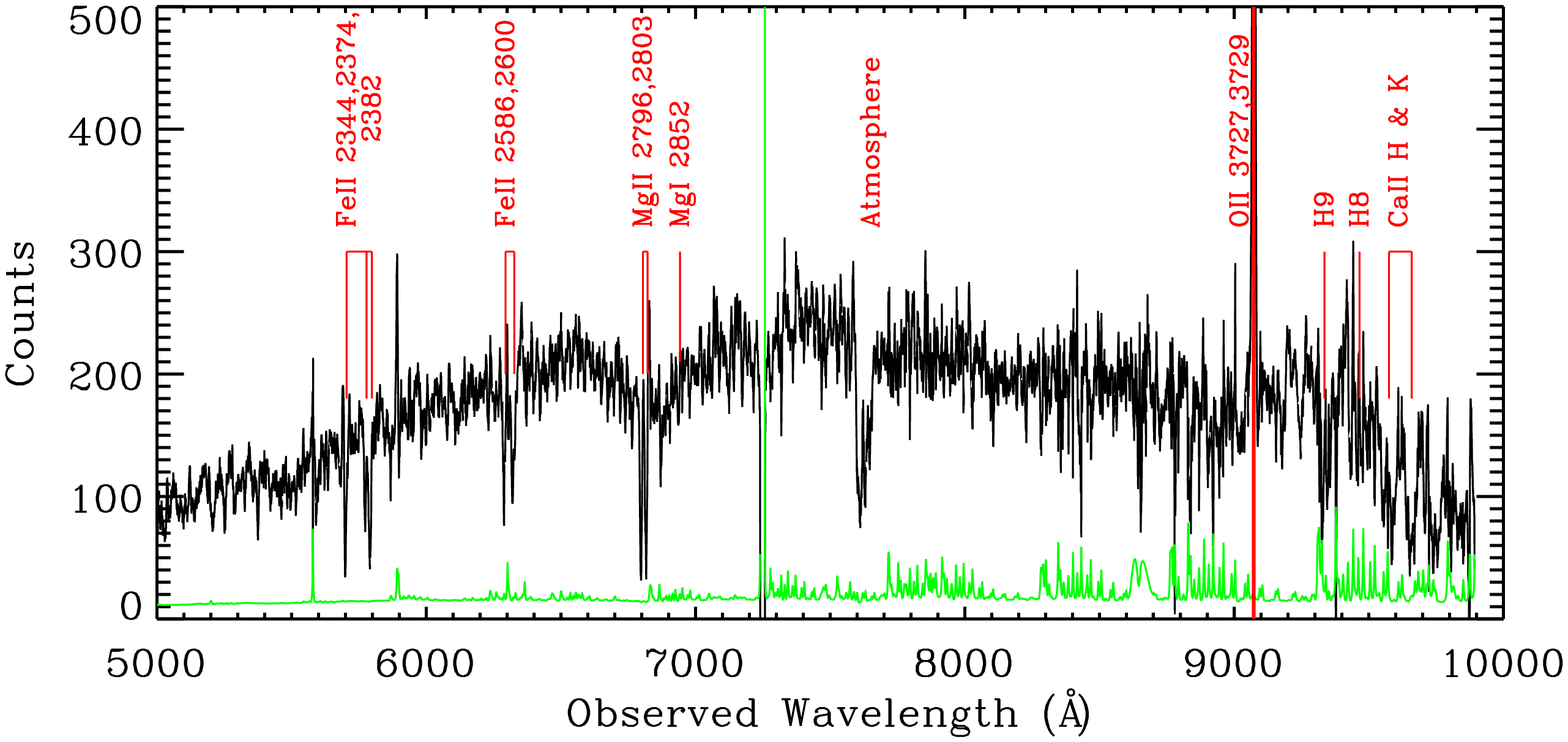}
\caption{HST/ACS color image of TKRS2158 in the $V_{606}$, $i_{775}$ and $z_{850}$ bands (left) and the TKRS spectrum inverse-variance smoothed using a window of 9 pixels (right).  The error in the smoothed spectrum is shown in green.  The absorption feature at $\sim 7650$ \AA \ is due to the atmosphere.  Absorption due to \ion{Mg}{2} and \ion{Fe}{2} is evident at observed wavelengths $\sim 5700 - 6900$ \AA.  The mean $\rm S/N$ near the \ion{Mg}{2} transition is $\rm 3.4~pix^{-1}$.  The object has an irregular morphology with several bright blue knots, suggestive of a recent or ongoing merger.  \label{fig.2158}}
\end{figure*}

\begin{figure}
\includegraphics[angle=90,width=3.25in]{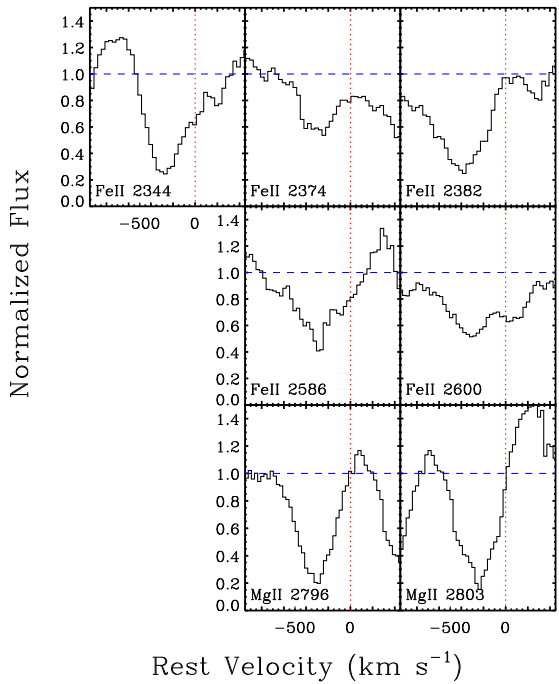}
\caption{
UV absorption lines in the spectrum of TKRS2158.  The spectrum has been inverse-variance smoothed using a window of 9 pixels.  The systemic velocity is marked with vertical dotted lines, and the dashed lines mark the continuum level.  All absorption lines shown are blueshifted by at least $-250~\rm km~s^{-1}$.  Narrow redshifted emission is evident in the MgII transition, resulting in a P Cygni-like line profile.  \label{fig.2158uvlinesstamps}}
\end{figure}

\begin{figure*}
\begin{center}
\includegraphics[angle=90,width=6.25in,viewport=30 0 700 490,clip]{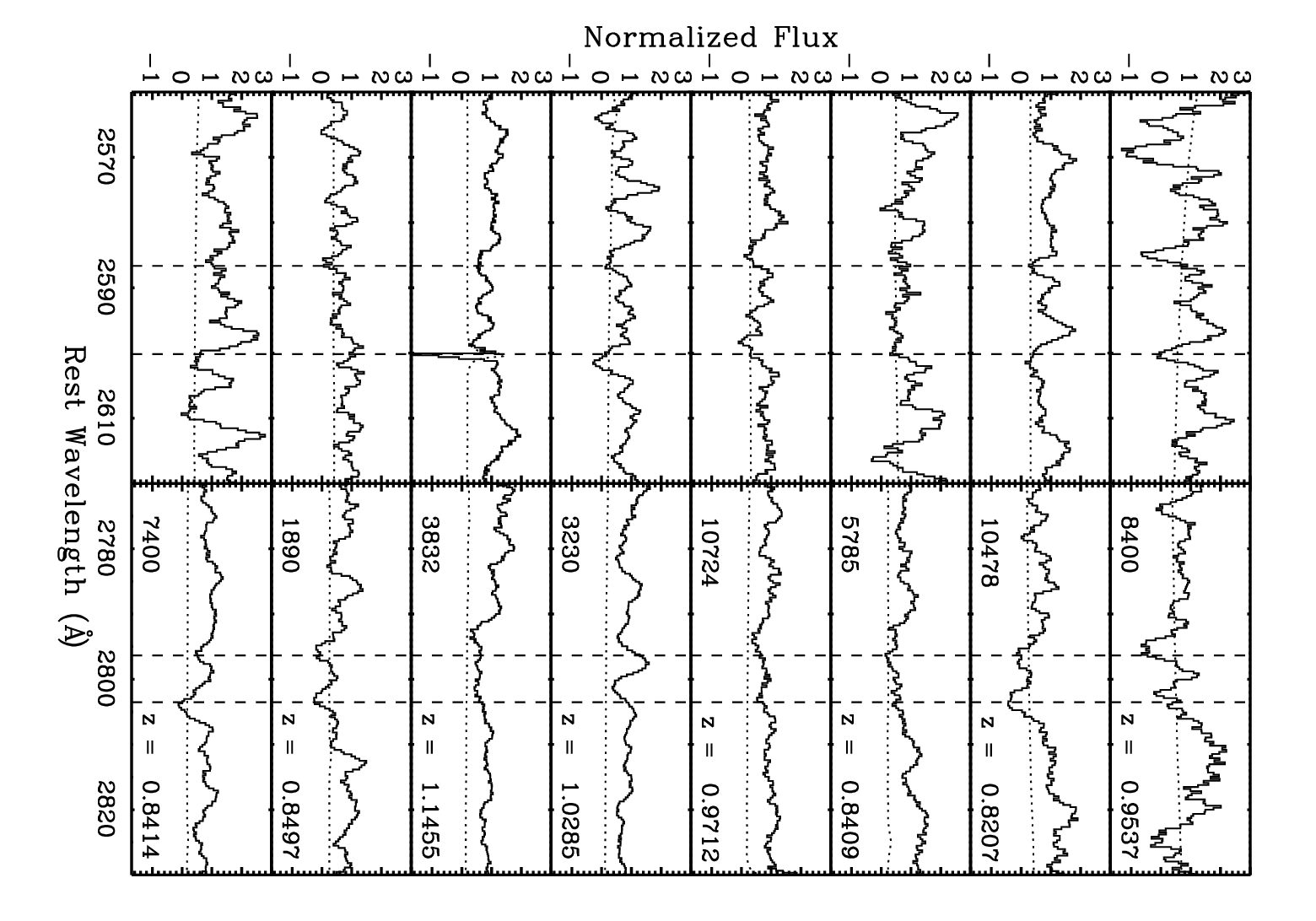}
\end{center}
\caption{UV absorption lines in TKRS spectra, inverse-variance smoothed using a 9 pixel window and normalized to the continuum level.  Objects are arranged from top to bottom in order of increasing apparent $R$ magnitude \citep{Wirth2004}, and the TKRS ID numbers are given in the right-hand panel.  The locations of the \ion{Fe}{2} $\lambda \lambda 2586, 2600$ and \ion{Mg}{2} $\lambda \lambda 2796, 2803$ transitions at the systemic velocity of each galaxy are marked with dashed lines.  The line profiles in the spectra of objects 8400, 10724, 3230, and 1890 are blueshifted with respect to systemic velocity, suggesting the presence of outflows.  Dotted lines show the error in each pixel.\label{fig.indiv_uvlines}}
\end{figure*}

One example is shown in Figure~\ref{fig.2158}, which includes a color HST/ACS image and the full spectrum of the object TKRS2158.  This object has a redshift $z = 1.43339$, has absolute $B$-band magnitude $M_B = -22.99$, and has an infrared luminosity $\rm \log L_{IR}/L_{\odot} = 11.9$ \citep{Melbourne2005}.    
Figure~\ref{fig.2158uvlinesstamps} includes sections of the same spectrum showing UV absorption line profiles.  The systemic velocity is given by the redshift of the galaxy reported in the TKRS, which was determined from the velocity of the [\ion{O}{2}] doublet.  Not only are \ion{Mg}{2} $\lambda \lambda 2796,2803$ and \ion{Fe}{2} $\lambda \lambda 2586,2600$ absorption lines detected, but \ion{Fe}{2} $\lambda \lambda 2344,2374,2382$ absorption is evident further to the blue.  
We calculate velocity offsets of each line with respect to systemic by fitting a Gaussian to each and finding the offset of its centroid.  These measurements are reported in Table~\ref{tab.veloff2158}.  
All lines are offset from the systemic velocity by $\sim -250 ~\rm km~s^{-1}$, with the exception of the \ion{Mg}{2} 2796 line and the \ion{Fe}{2} 2382 line, which are offset $-314 \pm 19 \rm~km~s^{-1}$ and $-415 \pm 36\rm~km~s^{-1}$ respectively.   These kinematics are a clear indication of cool outflow from this galaxy.  Discrepancies in the measurements of the velocity offsets could be due to noise in the spectrum such as poorly subtracted sky emission at the locations of certain absorption lines.  Additionally, narrow emission in the \ion{Mg}{2} 2796 transition or in \ion{Fe}{2}* fine-structure transitions may be filling in the some of the absorption profiles and shifting their centers to the blue \citep[e.g.,][]{Rubin2009}.

Figure~\ref{fig.indiv_uvlines} shows the spectral regions around the UV absorption lines for several more individual TKRS objects.  Blueshifted absorption lines are evident in a few of these spectra, hinting at the presence of outflows.  However, most of the spectra are not of sufficiently high quality to reliably measure velocity offsets or line strengths for individual galaxies.

\subsection{Outflow Signature in Coadded Spectra}

\begin{figure}
\includegraphics[width=3.6in,angle=90,viewport=20 -10 500 470,clip]{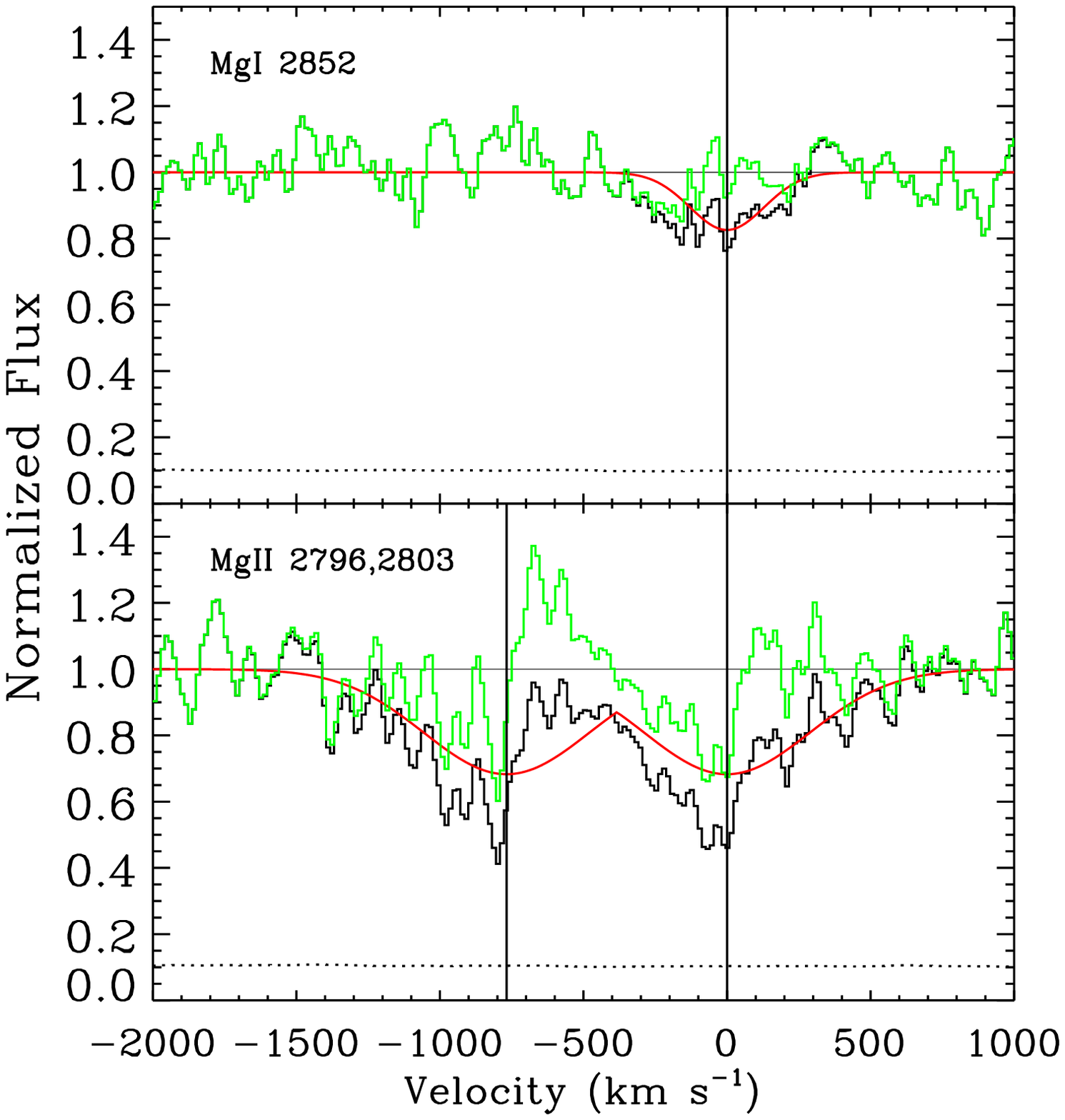}
\caption{Section of the coadd of all TKRS spectra around \ion{Mg}{2} and \ion{Mg}{1}.  The coadd
has been normalized to the level of the continuum surrounding the absorption
lines as described in the text and is plotted in black.  The dotted line is the error array for 
the coadd, and the black vertical lines mark the systemic velocity for each transition.  The red line overplotted on the coadd shows the Gaussian fits to the red sides of the $2803~\rm \AA$ and 2852 \AA \ lines; the Gaussian model profile for the systemic absorption in the 2803 \AA \ transition is 
assumed to describe the systemic absorption in the $2796~\rm \AA$ transition as well.
The green spectrum shows the absorption remaining after the models have been
divided out of the coadd. The black horizontal line marks the continuum level.  \label{fig.decomp}}
\end{figure}

Figure~\ref{fig.decomp} shows sections of the coadd of the full galaxy sample (\fullsampsize objects) surrounding the \ion{Mg}{2} and \ion{Mg}{1} $\lambda 2852$
absorption lines.  The spectrum has been normalized by a linear fit to the continuum flux in the continua regions $-2400 \rm ~km~s^{-1} < v < -1600~km~s^{-1}$ and $800 \rm ~km~s^{-1} < v < 1600~km~s^{-1}$, where $\rm v = 0~km~s^{-1}$ at $2803.531\rm~ \AA$ and $2852.964\rm~ \AA$ for \ion{Mg}{2} and \ion{Mg}{1} respectively.  
This continuum normalization is applied to all coadds presented.
The \ion{Mg}{2} line profiles in this figure are asymmetric, with more absorption on the blue side of $\rm v = 0~ km~s^{-1}$.  The \ion{Mg}{1} absorption line is asymmetric in the same sense.  
Quantitative analysis of these lines will be performed in later sections.

\begin{figure}
\includegraphics[width=4.8in,angle=90,viewport=40 -10 700 460,clip]{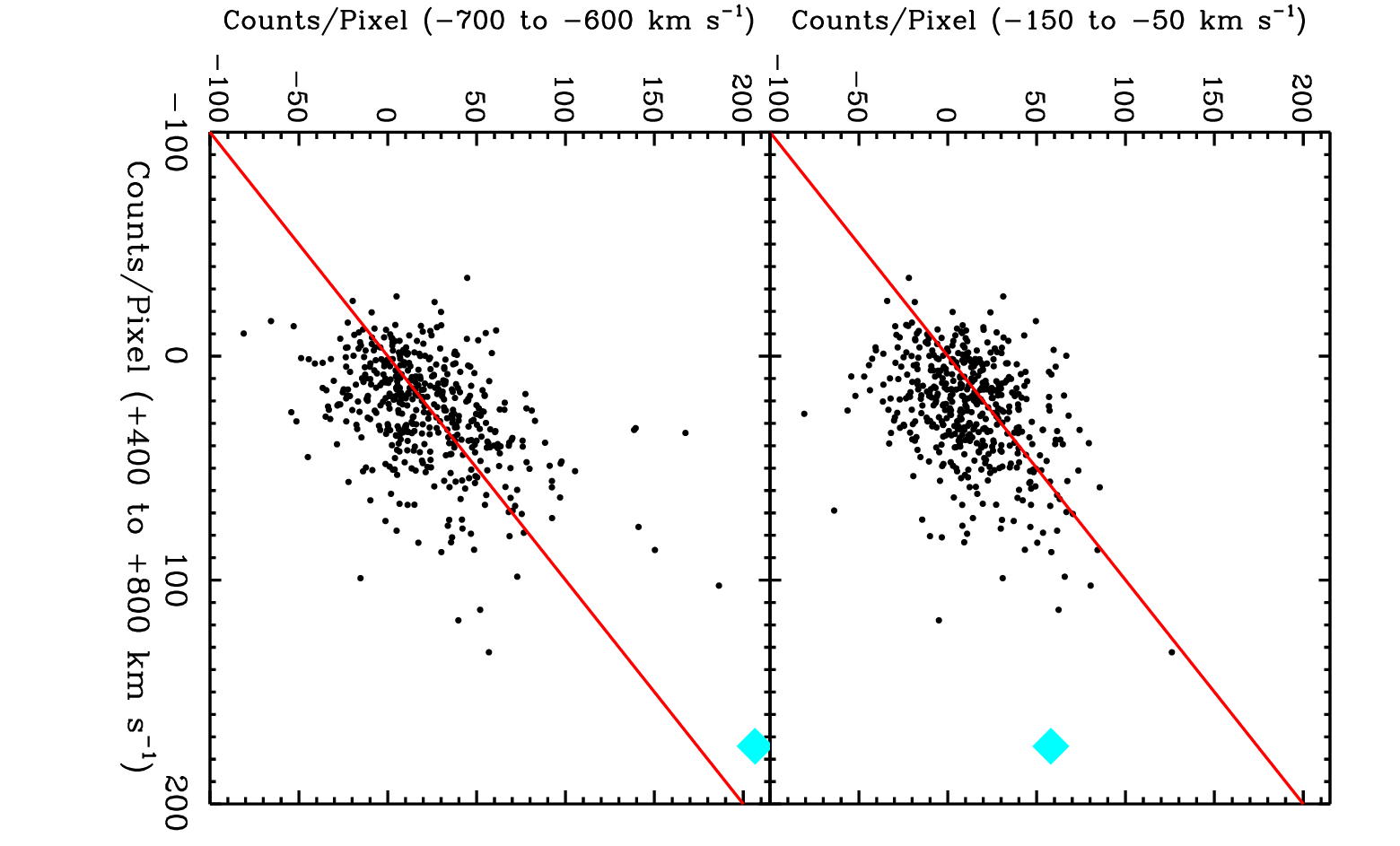}
\caption{\emph{Top:} Average $\rm counts/pixel$ in window 2 vs.\ window 1.  The points are offset to below a one-to-one ratio (shown in red), which indicates that a flux decrement to the blue of the $2803~\rm \AA$ \ line is a property of many of the galaxy spectra in the sample.  \emph{Bottom:} Average $\rm counts/pixel$ in window 3 vs. window 1.  Only a few objects are separated from the main locus with high values in window 3, indicating there are only a few objects with ``excess emission".  The location of the spectrum of TKRS2158 is marked in cyan in both panels.  
\label{fig.excessem}}
\end{figure}

To test whether the properties of the coadd reflect the properties of most of the individual galaxies in the sample, rather than a few of the brightest galaxies, we measure the flux in various velocity ranges with respect to \ion{Mg}{2} $\lambda 2803$ in individual spectra.  This was done in W09, and our velocity windows match those used in that work: $400~\rm km~s^{-1} < v < 800~\rm km~s^{-1}$ (window 1), and $-150~\rm km~s^{-1} < v < -50~\rm km~s^{-1}$ (window 2).  
Window 1 provides a measurement of the continuum and window 2 probes the flux in the deepest part of the blueshifted 2803 \AA \ line (although this velocity range may not probe blueshifted absorption in all spectra, due to errors in systemic velocity determinations as discussed in \S\ref{sec.stacking}).  We plot the average number of counts per pixel weighted by the inverse variance in window 2 vs. window 1 in the top panel of Figure~\ref{fig.excessem}.  The points are offset below a one-to-one ratio, showing that many of the spectra have a count decrement in window 2 (where blueshifted absorption is expected) relative to window 1 (where we are making a measurement of the counts in the continuum).  Approximately $2/3$ 
of the spectra have average counts lower in window 2 than in window  1, which indicates that a majority of the spectra contribute to the coadded flux decrement.  

We also wish to characterize the presence of emission in the \ion{Mg}{2} transition in individual galaxy spectra.  This is motivated by the work of W09, who identify \ion{Mg}{2} emission in coadded spectra as well as in $\sim 4$\% of the individual galaxy spectra in their sample at $z \sim 1.4$.  This emission may be due to AGN activity, but could also be related to some other physical process.  W09 attempt to exclude galaxies which exhibit \ion{Mg}{2} emission from their analysis of outflow properties in coadded spectra, as the complicated continuum shape of these individual spectra makes characterization of the \ion{Mg}{2} absorption line profiles in the coadds difficult.  To identify individual spectra with \ion{Mg}{2} emission in our sample, we measure the flux level in the velocity range $-700~\rm km~s^{-1} < v < -600~\rm km~s^{-1}$ (window 3) and compare it to our measurements in window 1 in the bottom panel of Figure~\ref{fig.excessem}. 
The placement of window 3 is immediately to the red of the $2796~\rm \AA$ line, and  
samples the region between the absorption lines where \ion{Mg}{2} emission is likely strongest.
Points toward the upper part of this plot have ``excess emission" in window 3.  There are 7 spectra in our sample with $\rm Counts/Pixel$ in window 3 greater than 120.  Four of these spectra have P Cygni-like line profiles at \ion{Mg}{2}; the three remaining spectra are pushed into the ``excess emission" regime because of noise in the window.  Removal of these spectra from our sample does not significantly affect the \ion{Mg}{2} line profiles in the coadded spectrum.  Their inclusion will therefore not demand careful modeling of emission in the continuum, and we retain them in the following analysis.

\section{Analysis of average outflow properties traced by \ion{Mg}{2} absorption}\label{sec.analysis}
In this section we introduce two measurements which will be used in the remainder of the paper  
 to assess the strength of the absorption due to outflowing gas in coadded spectra.  With both methods we attempt to estimate the strength of absorption at the systemic velocity through analysis of the red side of the \ion{Mg}{2} 2803 line.  We then ``correct" the measurement of the absorption strength on the blue side of the \ion{Mg}{2} 2796 line accordingly.  This corrected measurement should then depend on the absorption strength of outflowing gas only.
The first method uses boxcar EWs for the measurements in each velocity range and is useful in the case of low $\rm S/N$ coadds, while the second includes more detailed fitting of the line profile.  The latter method, however, requires that the red side of the \ion{Mg}{2} 2803 line can be well-characterized by a Gaussian.  

Analysis of the absorption strength of outflowing gas is complicated by absorption from stellar atmospheres and the ISM near the systemic velocity.
 Photospheric \ion{Mg}{2} absorption at the systemic velocity is strongest in F8 - G1 type stars \citep{Kinney1993}, in which the EW of the 2796 \AA \ line can reach $\sim 7$ \AA \ (see \S\ref{sec.feiimain}).
 As outlined in W09, 
 very bright late B, A and F stars
 can exhibit asymmetric or shifted \ion{Mg}{2} absorption due to stellar winds \citep{Snow1994}; these stellar
 absorption lines may be blueshifted by $\sim -100\rm~ km~s^{-1}$.  
 There may also be absorption due to the ISM of the
 galaxies; this gas is in the disks of the galaxies and will 
 rarely be blueshifted by more than a few tens of $\rm km~s^{-1}$  
 (i.e., in the case of a rotating disk).  
 The situation for \ion{Mg}{1} is simpler; stellar \ion{Mg}{1} absorption is not blueshifted and is only
 found in the photosphere of F stars.  It may be present in
 the ISM, however, and in outflowing gas.

 \subsection{Boxcar Method}
 In order to characterize the amount of absorption due to outflowing gas in each coadd,
 we make the assumption that the \ion{Mg}{2} doublet is saturated in our spectra, such that 
 the 2796 \AA \ and 2803 \AA \ lines have the same depths.  \ion{Mg}{2} becomes saturated at column densities ($N$) of $\rm \approx 10^{14}~ cm^{-2}$, which occur at 
 relatively low hydrogen column densities of $> 10^{19}~\rm cm^{-2}$ at solar abundance.  The ISM and stellar atmospheres typically have columns exceeding this value.  See \S\ref{sec.column} for further discussion.
 We measure $W_{diff}$, where 
  \begin{eqnarray}
  	W_{diff} &=& W_{2796 \rm\AA,blue} - W_{2803 \rm\AA,red}, 
  \end{eqnarray}
  and where
  \begin{eqnarray}
  	W_{2796 \rm\AA, blue} &=& \sum_{v = -500~\rm{km~s^{-1}}}^{0~\rm{km~s^{-1}}} \left [1 - \frac{f(v)}{f_c(v)} \right ] \Delta v;\\
	W_{2803 \rm\AA, red} &=& \sum_{v = 0~\rm{km~s^{-1}}}^{500~\rm{km~s^{-1}}} \left [1 - \frac{f(v)}{f_c(v)} \right ] \Delta v.
  \end{eqnarray}
  $W_{2803 \rm\AA,red}$ quantifies the amount of absorption due to gas that is not outflowing, and we subtract it from $W_{2796 \rm\AA,blue}$ to avoid overestimating the outflow absorption strength from the inclusion of absorption due to gas at $\rm v \approx 0~km~s^{-1}$ associated with the ISM and/or stellar atmospheres.

  \subsection{Decomposition Method}\label{sec.decomp}

 In addition to the boxcar measurement discussed above,   
 we also adopt the method of W09
 to first remove the stellar and ``stationary" ISM absorption
 from the line profile before making measurements of any outflow.  
 We use the model presented in W09:
 
 \begin{eqnarray}
  	F_{obs} (\lambda) &=& C(1 - A_{sym})(1 - A_{flow})\\
	A_{sym} &=& A_{2796} G(v,\lambda_{2796}, \sigma)  \nonumber \\
	& & {} + A_{2803} G(v,\lambda_{2803}, \sigma) \label{eq.gaussian}
  \end{eqnarray}
  
\noindent where $F_{obs} (\lambda)$ is the observed flux density, C is the galaxy continuum emission, and $A_{sym}$ and $A_{flow}$ are the line profiles of the symmetric and blueshifted (outflow) absorption. Here we make no attempt to include emission in \ion{Mg}{2} above the continuum in our model; this issue will be addressed in more detail in \S\ref{sec.emission}.  
$A_{sym}$ is made up of the sum of two Gaussians centered at the rest velocity of each line in the doublet.  In order to calculate $A_{sym}$, we fit
 a Gaussian profile to the red side ($\rm 0~km~s^{-1} < v < 1600~km~s^{-1}$) of the $2803~\rm \AA$ and $2852~\rm  \AA$ lines.  See
 Figure~\ref{fig.decomp} for a demonstration of this procedure.  Because
 the \ion{Mg}{2} doublet is blended in our spectra, we simply impose the Gaussian 
 fitted to the $2803~\rm \AA$ line onto the profile of the $2796~\rm \AA$ line.  The depths
 of these two Gaussians are kept the same ($A_{2796} = A_{2803}$), since the \ion{Mg}{2}
 lines are mostly saturated.  It may be that the $2796~\rm \AA$ line is in 
 reality slightly deeper than the $2803~\rm \AA$ line; in this case we will underestimate the
 strength of the systemic absorption in this line.
 We divide the coadd by this model, and the resulting blueshifted (or outflow) absorption line
 profile is plotted in Figure~\ref{fig.decomp} in green.  
 
 We presume that this profile 
 contains absorption only from gas that is outflowing.  However, if the doublet ratio for the stationary absorption is larger than one, we will measure too much outflowing gas (and the emission to the red of the $2796~\rm \AA$ line will be artificially weakened).  Redshifted emission in the $2803~\rm \AA$ line also affects the amplitude and width of the Gaussian we fit to the absorption line profile, and can cause an overestimate of the EW of outflowing gas.  Emission in the 2796 \AA \ line (evident in Figure~\ref{fig.decomp}) may in turn cause us to underestimate the EW of the outflow (see \S\ref{sec.emission} for more discussion of these effects).
 Finally, we are assuming that little of the absorbing gas is flowing into the galaxies; if we were to detect a substantial inflow, we would overestimate the amount of absorption at the systemic velocity due to deeper profiles on the red sides of the lines, and underestimate the EW of outflowing gas.
 
 As can be seen in Figure~\ref{fig.decomp}, no absorption is evident at positive relative velocities in the line profile shown in green, i.e., the ``outflow" profile.  Some emission occurs on the red side of the $2796~\rm \AA$ line; this will be discussed in 
 \S\ref{sec.emission}.  The EW 
 of the feature on the blue side of the $2796~\rm \AA$ line is $0.51 \pm 0.14 \rm~  \AA$
 (measured between $-1132~\rm km~s^{-1} < v < 0~km~s^{-1}$ using a boxcar sum).
 Although the green profile for \ion{Mg}{1} is suggestive of an outflow, the outflow absorption is not 
 detected, with a $3\sigma$ upper limit of $\sim 0.3$ \AA, possibly indicating that the outflow is dominated by gas densities lower than those needed to 
 contain a significant column in \ion{Mg}{1} \citep{Murray2007}.
 
 To compute the error on the outflow EW, we first generate 1000 realizations of the symmetric absorption profile by adding noise to the best-fit symmetric model generated in the above procedure.  The level of the added noise is determined by the noise in the coadded spectrum.  Each of these realizations is fit with a Gaussian, and a double Gaussian profile ($A_{sym}$) is created as above.  We then calculate the standard deviation of the values of these 1000 different models at each pixel, which is an estimate of the error introduced by the model fitting procedure as a function of velocity.  This error is combined in quadrature with the error in the coadded spectrum itself, producing an error at each pixel of the outflow profile ($A_{flow}$).  It is then straightforward to use this error to calculate the uncertainty in EW measurements of the outflow profile.

\subsection{Sensitivity of $W_{diff}$ and Outflow EW to Winds}\label{sec.sn} 
We now explore the extent to which these two quantification methods are sensitive to outflows with a range of physical parameters.  To do this, we generate a series of model \ion{Mg}{2} absorption line profiles, each with varying amounts of absorption at systemic velocity and offset to negative velocities.  We model both the systemic absorption and the outflowing absorption as single velocity components with Gaussian optical depth ($\tau$).  While cool outflowing gas likely consists  of multiple absorbing clouds at different velocities \citep[e.g.,][]{Martin2005}, we are coadding our data, and thus we expect that such features will be completely smoothed out.  Components are parametrized following \citet{Rupke2005b}, with variable \ion{Mg}{2} column density ($N$(\ion{Mg}{2})), covering fraction ($C_f$), Doppler parameter ($b_D$), and central wavelength ($\lambda_0$).  In all of our models, we choose $N$(\ion{Mg}{2})$ = 10^{14.9}~\rm cm^{-2}$ for the systemic component, such that the profile is completely saturated, and $N$(\ion{Mg}{2})$ = 10^{14}~\rm cm^{-2}$ for the outflowing component.  All components (systemic and outflowing) have $C_f = 0.5$, close to the 55\% absorption depth of the saturated \ion{Mg}{2} profiles in W09 (which in coadded spectra corresponds to the detection frequency of outflows times the $C_f$ of cool outflowing clouds).  Outflowing components are given relative velocities of -100, -200, and -300 \kmsns; we also create models with no outflowing component.  In each model, the systemic and outflowing components have the same $b_D$; however, we allow this parameter to have the values $b_D = 50$, 150, and 250 \kms in different models.  We therefore have a grid of $3 \times 4$ models in $b_D$ and outflow velocity space.

We smooth each of these models to the velocity resolution of the individual spectra, $\rm \sigma_{inst} = 1.4$ \AA \ \citep{Weiner2006}, adjusted to the rest-frame at the median redshift of the sample ($z = 0.94$).  A velocity resolution element in the coadded spectra is larger than this, due to uncertainties in the redshift determinations and wavelength solutions for the individual spectra (see \S\ref{sec.spectra}); however, the results are similar if we repeat the analysis using a velocity resolution that is $\rm 2\sigma_{inst}$.  After smoothing we add different levels of noise such that the resulting spectra have $\rm S/N = 3$, 6, and 9 $\rm pix^{-1}$, consistent with the range in $\rm S/N$ levels of coadded spectra we create in \S\ref{sec.division} (see Table~\ref{tab.ew}).  We generate 1000 realizations of each model at each $\rm S/N$ level, and then measure $W_{diff}$ and perform our decomposition analysis for each realization.  

\begin{figure}
\includegraphics[width=4.75in,angle=90,viewport=30 -50 650 440,clip]{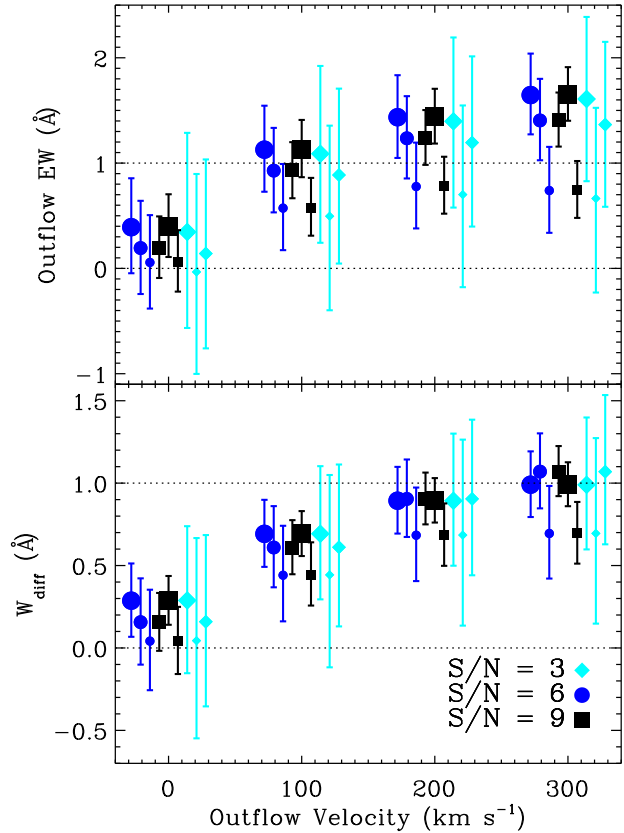}
\caption{\emph{Top:} Distribution in measured outflow EW values for model spectra as a function of the input outflow velocity.  The points are offset slightly in velocity to prevent overlap.  The mean outflow EW for realizations with $\rm S/N = 3~pix^{-1}$, $\rm 6~pix^{-1}$ and $\rm 9~pix^{-1}$ are shown with cyan diamonds, blue circles, and black squares respectively.  Small, medium, and large symbols show results for models with $b_D = 50$, 150 and 250 \kmsns.  The error bars indicate the 90\% confidence intervals for each set of model realizations.  \emph{Bottom:}  Same as above, for $W_{diff}$. 
\label{fig.sn}} 
\end{figure}

Figure~\ref{fig.sn} shows the distribution in the measured outflow EW and $W_{diff}$ values for all models as a function of the model outflow velocity.  The points have been offset by an arbitrary amount in velocity so that they do not overlap.  The mean outflow EW and $W_{diff}$ for models with $\rm S/N = 3~pix^{-1}$, $\rm 6~pix^{-1}$ and $\rm 9~pix^{-1}$ are shown with cyan diamonds, blue circles, and black squares respectively.  The point size increases with $b_D$.  The error bars show the 90\% confidence intervals in the measured quantities.  The size of these intervals and the central values indicate the degree to which outflow EW and $W_{diff}$ are successful in characterizing the underlying physical absorption profile.
For instance, at $\rm S/N = 3~pix^{-1}$, $W_{diff}$ and outflow EW have very broad distributions for all models.  
For models with no outflow, both of these measures exceed values of 0.5 \AA \ in between 12\% and 40\% of realizations, with outflow EWs for models with $b_D = 150\rm ~km~s^{-1}$ and $250\rm ~km~s^{-1}$ exceeding $\sim 1$ \AA \ in 6\% and 12\% of realizations, respectively.  Likewise, models with outflows can have $W_{diff}$ or outflow EW $< 0.5$ \AA, although in general the distributions shift to higher values of these quantities as outflow velocities increase.  We conclude, however, that in the case of coadds with $\rm S/N \sim 3~pix^{-1}$, we are able to confirm the presence of outflows only if the measured values of outflow EW are $\gtrsim 1.2-1.3$ \AA \ and $W_{diff}$ values are $\gtrsim 0.8$ \AA.  These values are recovered for outflow velocities $\gtrsim 100 \mkms$, and occur more frequently with higher $b_D$.  If lower values are measured, the presence or lack of outflows remains ambiguous.  

These measurements become more sensitive with increasing $\rm S/N$.  At $\rm S/N = 6~pix^{-1}$, $W_{diff}$ and outflow EW can be as high as 0.5 - 0.9 \AA \ for models with no outflow.  However, models with outflows yield mean values higher than these if $b_D = 150$ or 250 \kmsns.  Values of outflow EW $> 0.85$ \AA \ are measured for at least 95\% of realizations of models with high values of $b_D$ and outflow velocity.  At $\rm S/N = 9~pix^{-1}$, models with no outflow have $W_{diff}$ and outflow EW $\lesssim 0.4 - 0.7$ \AA, while models with outflows and $b_D = 150$ or 250 \kms again yield higher values in at least 92\% of realizations.  Models with $b_D = 50$ \kms have higher mean values if the outflow velocity $ \gtrsim 200$ \kmsns.

Although this modeling relies on a very simplistic parameterization of the \ion{Mg}{2} absorption profiles in our coadded spectra, it suggests that we are sensitive to saturated \ion{Mg}{2}-absorbing outflows with velocities on the order of 100 \kms in coadds with $\rm S/N \sim 6-9~pix^{-1}$.  We are sensitive to outflows in $\rm S/N \sim 3~pix^{-1}$ coadds only with large velocities and $b_D$.  These findings will be discussed further in \S\ref{sec.discussion}.

\subsection{\ion{Mg}{2} in Emission}\label{sec.emission}
We observe emission in the \ion{Mg}{2} transition in both individual galaxies (see the P-Cygni profile in Figure~\ref{fig.2158uvlinesstamps}) and in coadded spectra (e.g., Figure~\ref{fig.decomp}).  Emission in the latter is obvious after the decomposition of systemic and outflow profiles is performed and one can identify the decrement of absorption on the red side of the 2796 \AA \ line as compared to the red side of the 2803 \AA \ transition.  This emission is observed in $z \sim 1.4$ star-forming galaxies (W09), a luminous starburst galaxy at $z \sim 0.7$ \citep{Rubin2009}, as well as in narrow-line Seyfert galaxies such as two of the ultraluminous infrared galaxies (ULIRGs) studied in \citet{MartinBouche2009}.  A similar P-Cygni-like profile in the \ion{Na}{1} transition was observed in NGC 1808, a starburst galaxy driving an outflow \citep{Phillips1993}.  While the origin of this emission remains unclear, we suggest it may be at least in part due to resonance-line scattering off of the receding side of an expanding shell related to the observed outflow, as in the case of Ly$\alpha$ emission in LBGs \citep[][K. Rubin et al., 2009, in preparation]{Pettini2001}.  

To better understand the effect of this emission on our outflow EW measurements, we develop a alternative method of characterizing the \ion{Mg}{2} doublet profiles in coadded spectra.  We again assume that the absorption at the systemic velocity produces a saturated Gaussian absorption profile as in Equation~\ref{eq.gaussian}.  We then further assume that there is additional emission on top of this continuum on the red side of each line with a Gaussian profile.  The amplitudes of the emission lines have a ratio 2:1, and the lines have a variable velocity offset with respect to systemic.  In a true P-Cygni profile, the velocity at the peak of the emission (and the shape of the profile in general) depends on a number of factors, e.g.,\ the outflow geometry and $C_f$, and the velocity dispersion of the outflowing gas.  Our model can be written as follows:

 \begin{eqnarray}
  	F_{obs} (\lambda) &=& C(1 - A_{sym})(1 - A_{flow}) + F_{em}\\
	F_{em} (\lambda)  &=& A_{2796} G(v-v_0,\lambda_{2796}, \sigma) \nonumber \\
	& & {} + A_{2803} G(v-v_0,\lambda_{2803}, \sigma) 
  \end{eqnarray} 

\noindent where $F_{em}$ is emission in excess of the continuum and $ A_{2796} = 2~ A_{2803}$.  We may then fit this model to the red sides of \emph{both} lines in the doublet simultaneously (where $A_{flow} = 0$), in the velocity ranges 
$\rm 0~km~s^{-1} < v < 268~km~s^{-1}$ and $\rm 0~km~s^{-1} < v < 1600~km~s^{-1}$ for the 2796 \AA \ and 2803 \AA \ lines, respectively.  
We subtract the fitted emission model ($F_{em}$) from the coadd, divide out the model symmetric absorption, and measure the boxcar EW in the range $-1132~\rm km~s^{-1} < v < 0~km~s^{-1}$ for the resulting profile.

Because of the number of free parameters in this model (5), this method does not generally produce acceptable fits for the coadded spectra in our study, as the results are often driven by noise features in the line profiles.  However, the model does successfully characterize the red sides of the doublet lines in the much higher-$\rm S/N$ coadds of W09.  We find that in the W09 coadds with the strongest emission features (e.g., the lowest- and middle-$\rm M_*$ subsamples), the outflow EW we calculate using this method is $\sim 6 - 17$\% lower than the outflow EW calculated in W09 (and described above in \S\ref{sec.decomp}).  In coadds such as the high-$\rm M_*$ W09 subsample, the difference in outflow EWs calculated with the two methods is $< 3$\%.  While this model is quite simplistic, these results indicate that \ion{Mg}{2} emission causes an overestimate of the outflow EW calculated using our standard method (\S\ref{sec.decomp}).  A more complete, physically-motivated model including radiative transfer is required to fully quantify this effect, and additionally may more tightly constrain other characteristics (e.g.,\ radial extent, density) of the cool gas outflow.

\section{Properties of sample galaxies}
We wish to characterize outflow absorption strength at $0.7 < z < 1.5$ in galaxies with a range in SFR, $\rm M_*$ and $\rm \Sigma_{SFR}$ as well as explore the evolution of outflows.  To do this, we create subsamples of the galaxies with similar SFRs, $\rm M_*$s, etc., and coadd spectra within a given subsample.  We then compare outflow properties among these coadds.  Here we describe how photometry, sizes, $\rm M_*$, SFR, and quantitative morphologies are derived for our sample.

 \subsection{Rest-frame Colors and Luminosities}
 We use photometry from \citet{Weiner2006} derived from ACS imaging \citep{Giavalisco2004} and \citet{Capak2004} ground-based photometry, and converted to absolute $M_{B}$ and rest-frame $U-B$ color using the K-correction routine of \citet{Willmer2006}.  
 Errors in the observed optical magnitudes and colors are 0.05-0.07 mag and 0.07-0.1 mag, respectively. 
 The $1\sigma$ errors introduced by the K-correction procedure are 0.12 for $M_B$ and 0.09 in $U-B$ \citep{Weiner2006}.  
 The left-hand side of Figure~\ref{fig.cmd_masssfr} shows a color-magnitude diagram (CMD) for our sample.  The solid line is from \citet{Willmer2006}, and marks the division between the red sequence, or the narrow region populated by early-type E/S0s in the CMD, and the blue cloud, or the wider area in the CMD populated by bluer spirals and separated from the red sequence by a narrow ``valley" in the surface density of objects.  
 
 \subsection{Sizes}
 We wish to measure $\rm \Sigma_{SFR}$ for our sample, where $\rm \Sigma_{SFR} = SFR / \pi R_{SF,1/2}^2$, for comparison with the suggested local threshold for driving outflows, \thresh.  In the local Universe, measurements of the SFR from H$\alpha$, far-IR or extinction-corrected UV fluxes are combined with measurements of the sizes of star-forming regions ($\rm R_{SF,1/2}$) to determine $\rm \Sigma_{SFR}$ \citep[e.g.,][]{Meurer1997,LehnertHeckman1995}.  The sizes are constrained using measurements of the half-light radius from H$\alpha$ or UV (e.g., 2200 \AA ) imaging, which are direct tracers of nebular emission or emission from young stars.  Local starburst galaxies with high values of SFR per unit star-forming surface area (several $\rm M_{\odot}~yr^{-1}~kpc^{-2}$) have values of SFR per unit surface area in the optical disk, i.e., $\rm SFR / \pi D_{25}^2$\footnote{$\rm D_{25}$ is the apparent isophotal diameter measured at surface brightness $\rm \mu_B = 25~mag/\square \arcsec$}, which are at least 3 orders of magnitude lower \citep{Martin1999}.  It is therefore important to probe the spatial extent of UV flux when making size measurements of distant galaxies for comparison with local results.  Continuum emission at redder wavelengths arises from older stellar populations, which are not expected to be important for driving large-scale outflows.
 
We use the half-light radii of the TKRS galaxies measured by \citet{Melbourne2007} to parametrize galaxy size.  These authors fitted successively larger elliptical apertures to the ACS images for each galaxy and calculated the fluxes and intensities within each aperture.  An iterative curve-of-growth analysis was used to determine the flux level of the sky.  From these measurements, the total flux of the object was calculated.  Apparent half-light radii are equal to the semimajor axis of the ellipse that contained half of the total flux, and were corrected for the point-spread function of the image.  
To determine half-light radii at 2200 \AA \ in the rest-frame, we interpolate between the radii measured in the passbands to the red and blue of 2200 \AA \ in the rest-frame of each object, assuming that all of the light measured in each band is observed at the central wavelength of the filter (4297 \AA \ for the $B_{435}$ band and 5907 \AA \ for the $V_{606}$ band).  In cases in which 2200 \AA \ in the rest-frame is blueward of the $B_{435}$-band filter, we simply adopt the $B_{435}$-band radius.  We then use the angular diameter distance to compute the rest-frame UV half-light radius in kiloparsecs ($\rm R_{1/2}$).  \citet{Melbourne2007} estimate that they obtain accurate radii to within $< 10$\%.  This level of uncertainty applies strictly to the rest-frame $B$-band half-light radii they derive by combining the radii measured in the observed bands in a weighted mean, with weights dependent on the overlap of each observed passband with the rest-frame $B$-band.  We assume this level of uncertainty applies to our $\rm R_{1/2}$ as well.
 
We choose to use $\rm R_{1/2}$ to parametrize the sizes of the star-forming regions in our galaxies for its simplicity; however, this measure may in fact significantly overestimate the size scales relevant for driving outflows.  Many of the galaxies have extended and clumpy morphologies, such that $\rm R_{1/2}$ is quite large ($\rm > 10~ kpc$), while much of the UV emission arises in a few small but widely separated bright knots.  On the other hand, the distance between star-forming knots may be intimately connected to the morphology of the outflowing gas itself \citep[see, e.g.,][for some discussion of outflow morphology]{Heckman1990,Martin2006}.  Future studies of outflows in individual galaxies will warrant more careful analysis of the size scale and distribution of star formation.
 
 \subsection{Stellar Mass}
 Near-IR photometry of the GOODS-N field was published by \citet{Bundy2005}, who derived $\rm M_*$ for 202 objects in our sample (out of a total of \fullsampsize).  As we did not wish to limit this study to only objects with K-band derived $\rm M_*$, we use a calibration derived in W09 to convert rest-frame color and magnitudes into $\rm M_*$.  \citet{Bell2003} compute the relation between rest-frame color and $\rm M_{*}/L_B$ using SDSS and 2MASS photometry of local galaxies.  They give this relation for a ``diet Salpeter" IMF:
  \begin{eqnarray}
  	\log M_{*} / L_B (z = 0) &=& -0.942 + 1.737~ (B-V_{Vega}). \nonumber
  \end{eqnarray}
 This must be adjusted according to the redshift of each galaxy.  W09 derive a redshift correction to this relation using the $K$-band magnitudes and $\rm M_*$ (with Chabrier IMF) available for 11924 objects in the DEEP2 redshift survey \citep{Davis2003, Bundy2006}.  These objects lie in a redshift range $0 < z < 1.5$.  W09 performed a least-squares fit between $\rm M_*$ derived from rest-frame color and those derived from $K$-band photometry, and give a correction term $C_K$:
 \begin{eqnarray}
  	M_{*} &=& L_{B, Vega} \times M_{*}/L_{B} (z=0) \nonumber \\
	& & {} \times C_{K} (U-B, z); \nonumber \\
	\log~C_{K} (U-B, z) &=& -0.0244 - 0.398 z  \nonumber \\
	& & {} + 0.105~ (U-B_{Vega}). \nonumber
  \end{eqnarray}
 The authors find a scatter of 0.25 dex about the fit for $C_K$.  We apply this correction to our data, and compare $K$-band $\rm M_*$ to the corrected color-derived masses where possible (for 202 objects).  We find there is a 0.09 dex mean offset and a dispersion of 0.22 dex between the two stellar mass estimates. 
 We use the corrected color-derived masses with a Chabrier IMF for the full sample in the following analysis.

 \subsection{Star Formation Rate}\label{sec.sfr}
 The rich multi-wavelength data set in the GOODS-N field makes several different methods available for the derivation of the total SFR for the galaxies in our sample.  [\ion{O}{2}] line luminosities have been measured for TKRS galaxies \citep{Weiner2006}.  24 $\mu$m fluxes and IR-based SFRs for a subset of objects in our sample are available from \citet{Melbourne2005}.  In order to make comparisons with W09, we adopt their method of determining SFR.  The $BV$ photometry of \citet{Capak2004} measures the flux at 1800 \AA \ - 2800 \AA \ and 2200 \AA \ - 3400 \AA \ in the rest-frame for our redshift range.  We use these measurements to derive absolute magnitudes at $1500$ \AA \ and $2200$ \AA \ using the K-correction code described in \citet{Willmer2006} and \citet{Weiner2005}.  Using these luminosities we find the slope of the UV continuum, $\beta$, and calculate the attenuation from this slope using the relation $A_{FUV} = 3.16 + 1.69\beta$ \citep{Seibert2005,Treyer2007}.  There is a $\pm 0.9~$mag scatter in this relation, which results in a 0.36 dex uncertainty in the SFRs.  We use the values of $A_{FUV}$ to calculate an unextincted UV luminosity, and in turn calculate the SFR \citep{Kennicutt1998} for a Kroupa IMF:
  \begin{eqnarray}
  	\rm SFR_{UV} (M_{\odot}~ yr^{-1}) &=& 1.0 \times 10^{-28} L_{1500\rm \AA} \rm (erg~ s^{-1} Hz^{-1}). \nonumber
  \end{eqnarray}

To investigate systematic errors in these SFR estimates, we compare them to SFRs derived from the IR luminosities of \citet{Melbourne2005}.  These authors used the publicly available MIPS imaging of the GOODS-N field to measure 24 $\mu$m fluxes to a limit of 25 $\mu$Jy, and used the \citet{LeFloch2005} prescription to convert 24$\mu$m flux to total IR luminosity ($L_{IR}$).  \gdlir of the \bcsampsize blue cloud galaxies in our sample have $L_{IR}$ estimates available.  We use the \citet{Bell2005} relation between $L_{IR}$ and SFR with a Kroupa IMF to calculate SFR(IR) for these galaxies, ignoring the contribution from unextincted light from young stars in the UV, which we expect to be small \citep[][W09]{Bell2005}.  We find that for IR-selected galaxies, which may have slightly larger IR-to-UV emission ratios than is typical (W09), $L_{IR}$-derived SFRs are 0.21 dex higher than the UV-derived SFRs, with a dispersion of 0.43 dex.  
The expected offset will likely be smaller for galaxies not detected in the MIPS imaging.  For consistency with W09, and because we lack $L_{IR}$ measurements for nearly half of the blue cloud galaxies in our sample, we adopt the UV-derived SFRs described above in our analysis.  However, these uncertainties must be considered when absolute SFRs are discussed, as in \S\ref{sec.sfrsd}.  \citet{Bell2005} report systematic and random errors in their IR / UV derived SFRs of 0.3 and 0.4 dex; we must account for these errors in addition to those introduced from using purely UV-derived SFRs.

\begin{figure}
\includegraphics[width=3.5in,angle=90,viewport=-20 10 480 500,clip]{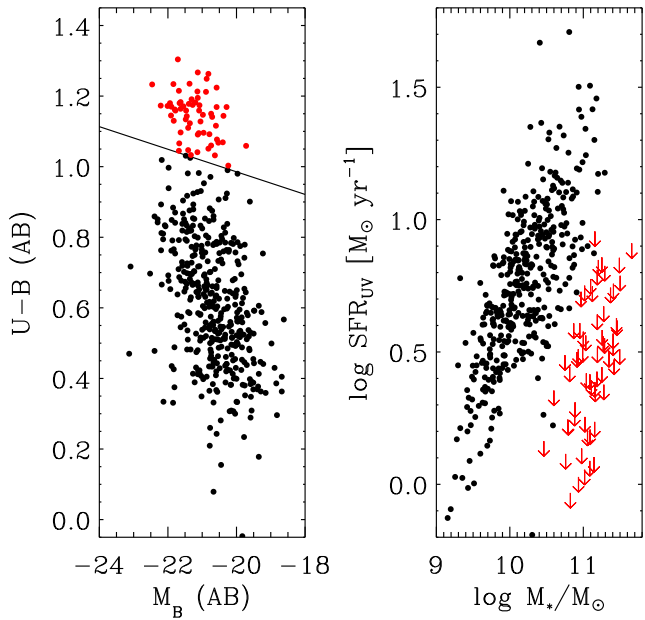}
\caption{\emph{Left:} Color-magnitude diagram for all objects in our sample.  The solid line marks the division between the red sequence and the blue cloud objects (in red and black, respectively); it is taken from \citet{Willmer2006}.  \emph{Right:} $\rm SFR_{UV}$ vs.\ $\rm M_{*}$ for the full sample.  Downward red arrows mark upper limits on the $\rm SFR_{UV}$ for objects on the red sequence.     \label{fig.cmd_masssfr}}
\end{figure}

 The right-hand side of Figure~\ref{fig.cmd_masssfr} shows a plot of $\rm \log SFR_{UV}$ vs.\ $\rm \log M_{*}$ for all objects in our sample with UV SFR measurements.  Upper limits on the SFRs in red sequence galaxies are marked in red.  While our sample includes galaxies in a range of $\rm M_*$ comparable to the sample of W09, the mean SFR in W09 is $\sim 0.57$ dex higher than in this study.  
 
 \subsection{Morphology}
\begin{figure}
\includegraphics[width=3.25in,angle=90]{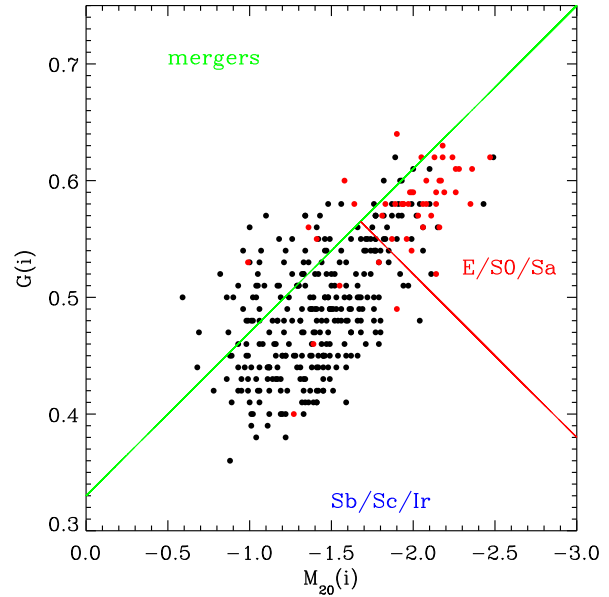}
\caption{G vs.\ $\rm M_{20}$ measured in the $i_{775}$-band for objects in our sample.  Measurements were provided by J. Lotz (2008, private communication).  The colored lines show the regions occupied by different morphological types and are adopted from \citet{Lotz2008}.  Red sequence galaxies are marked in red.  \label{fig.GM20}}
\end{figure}

 We use the Gini (G) and $\rm M_{20}$ measurements made for TKRS galaxies in the $i_{775}$-band by J. Lotz (2008, private communication) to quantify galaxy morphology.  These are nonparametric measurements described in \citet{Lotz2004,Lotz2006}.  G quantifies the relative distribution of light among a galaxy's pixels, and is high if there are only a few very bright pixels.  $\rm M_{20}$ is the second-order moment of the brightest 20\% of a galaxy's pixels; 
it is larger in galaxies in which the brightest pixels are furthest from each other.  Figure~\ref{fig.GM20} shows the distribution of these parameters for our sample (excluding 115 which lack high-quality $\rm G/M20$ measurements).    
The dividing lines for different morphological types are taken from \citet{Lotz2008}, who applied these divisions to galaxies at $0.2 < z < 1.2$.  
There are 63 mergers, 226 late-type objects, and 64 early-type objects in our sample.

\section{Division of sample by galaxy properties}\label{sec.division}
\subsection{Division by $\rm SFR_{UV}$, $\rm M_*$, and Morphology}
To examine trends in $W_{diff}$ (where $W_{diff} = W_{2796 \rm\AA,blue} - W_{2803 \rm\AA,red}$) with $\rm SFR_{UV}$, $\rm M_{*}$, and morphology, we have divided our spectra
into several different subsamples and coadded them.  We calculate $W_{diff}$ for each subsample.  These subsamples and measurements are listed in Table 
\ref{tab.ew}.  

First, we coadd only galaxies on the red sequence.    
The coadd has a very low $\rm S/N = 1.4~pix^{-1}$ in the continuum surrounding \ion{Mg}{2}, and so cannot be used to examine outflows in these objects.  Because of their low $\rm S/N$, we exclude all red sequence galaxy spectra from the subsamples described in the following, with the exception of the morphologically divided subsamples.

\begin{figure*}
\includegraphics[width=3.5in]{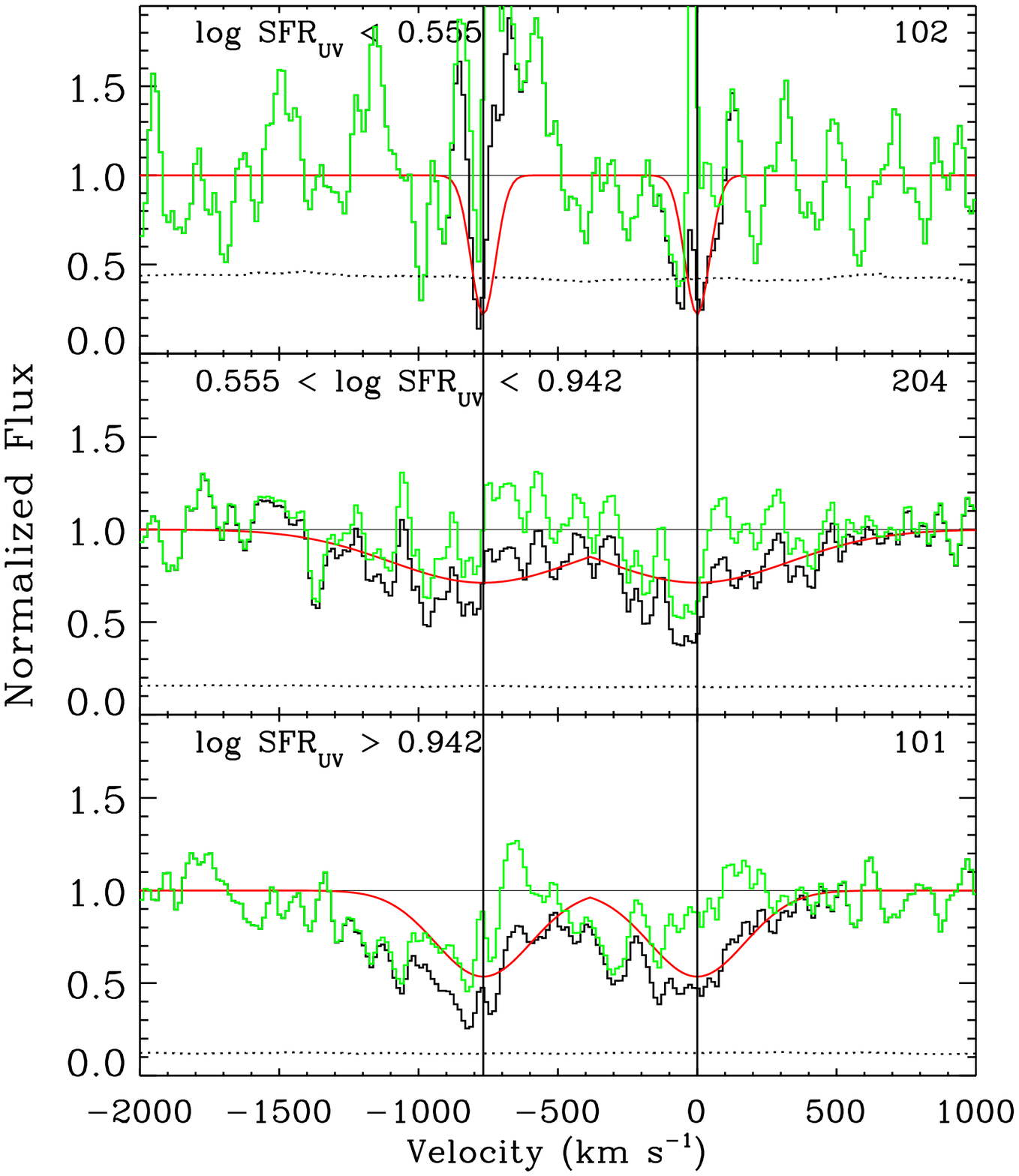}
\includegraphics[width=3.5in]{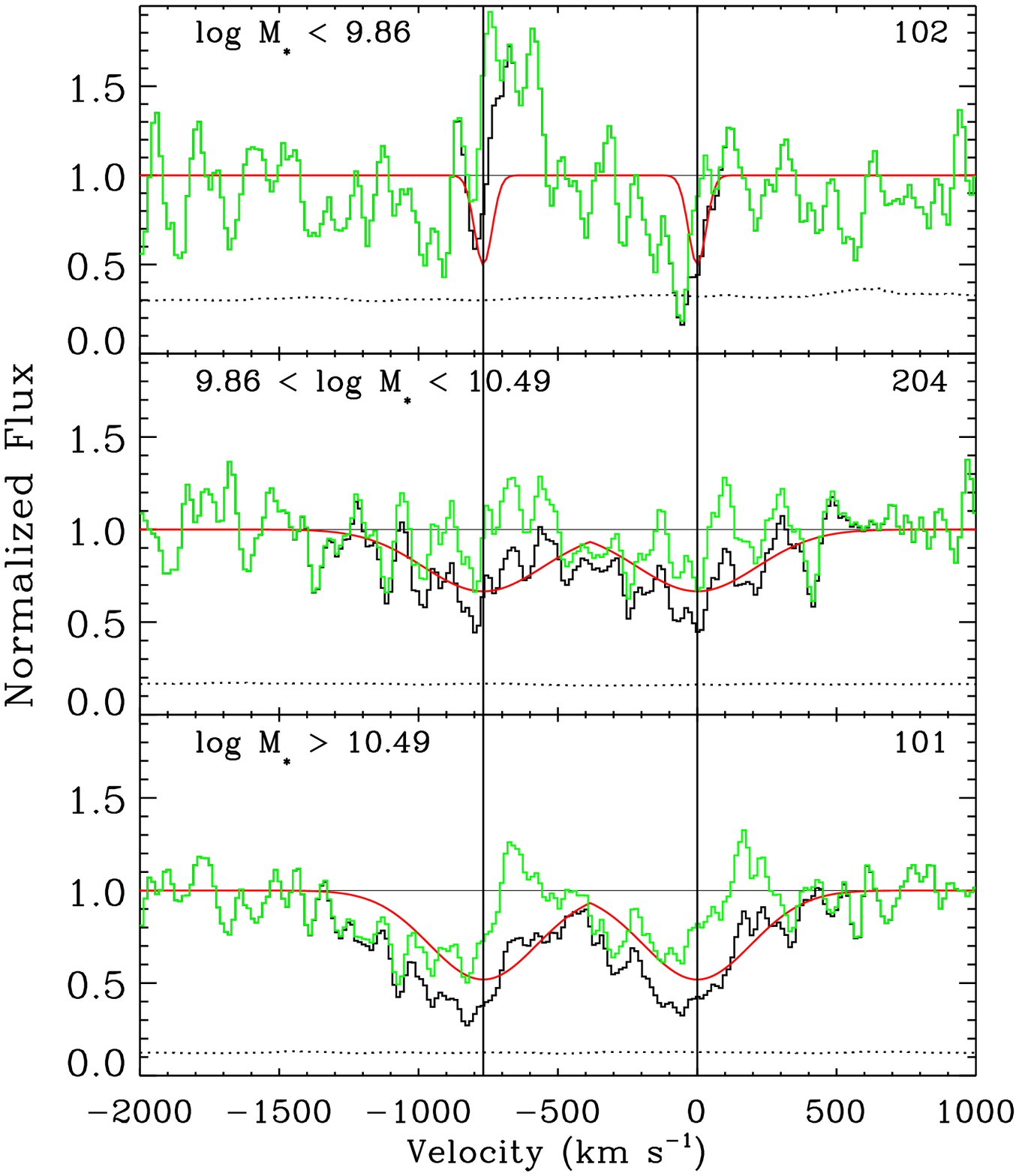}
\caption{Sections of the coadded spectra for subsamples divided by $\rm SFR_{UV}$ (left) and $\rm M_*$ (right) around \ion{Mg}{2} (black).    
The coadds
have been normalized to the level of the continuum surrounding the absorption
lines as described in the text.  The symmetric absorption profile is shown in red; the outflow profile is shown in green.  The error in each pixel is shown with the dotted lines, and the black vertical lines mark the systemic velocity for each transition.  The number of spectra in each subsample is shown in the upper right of each panel.  \label{fig.decompsfrspec}}
\end{figure*}

\begin{figure}
\includegraphics[width=3.5in,angle=90]{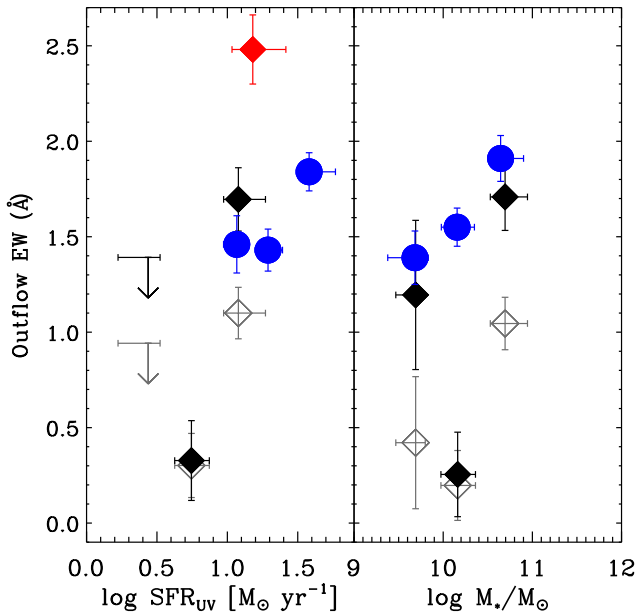}
\caption{Comparison between the outflow absorption strength, $\rm SFR_{UV}$ (left) and $\rm M_{*}$ (right).  Each point corresponds to a single coadd.  Gray open diamonds plot $W_{diff}$ measurements for the TKRS subsamples divided by the 25th and 75th percentile values of $\rm SFR_{UV}$ and $\rm M_{*}$.  Black diamonds show the outflow EW results from the decomposition procedure for the same subsamples.  The red diamond shows the outflow EW (from the decomposition procedure) for the ``DEEP2-like" subsample selected by $\rm M_{*}$ and $\rm SFR_{UV}$.  Blue circles show results from W09.  $2\sigma$ upper limits are shown when the error on the outflow absorption measurement ($W_{diff}$ or outflow EW) is twice the central value.
\label{fig.bxcr_outflow_SFR_M}}

\end{figure}

We choose to divide the TKRS galaxies by the 25th and 75th- percentile values of $\rm \log SFR$ (0.555, 0.942) and $\rm \log M_*$ (9.86, 10.49).  Coadds of the spectra for these subsamples are shown in Figure~\ref{fig.decompsfrspec}.  
Figure~\ref{fig.bxcr_outflow_SFR_M} shows $W_{diff}$ for the $\rm SFR_{UV}$-divided and $\rm  M_*$-divided subsamples.  The outflow absorption strength (as quantified by $W_{diff}$) rises significantly with $\rm SFR_{UV}$ and $\rm M_*$ between the middle and highest $\rm SFR_{UV}$ and $\rm M_*$ subsamples.   
There appears to be detected outflowing gas in the lowest-$\rm M_*$ subsample; however, this coadd has $\rm S/N \sim 3~pix^{-1}$, and thus is unlikely to yield reliable measurements (see \S\ref{sec.sn} for a discussion of this issue).
We also perform our decomposition analysis on each of these coadds, shown in Figure~\ref{fig.decompsfrspec}.  Measurements of outflow EW for each coadd are shown in Figure~\ref{fig.bxcr_outflow_SFR_M}, and show consistency with the $W_{diff}$ measurements.  Outflow EW results from W09 are shown in blue.  
Note that measurements of outflow absorption strength are significantly higher for the W09 $z \sim 1.4$ galaxies than for the middle-$\rm M_*$ TKRS galaxies.  As discussed in \S\ref{sec.sfr}, while the TKRS and W09 galaxies have a similar range in $\rm M_*$, on the whole the TKRS galaxies have a mean $\rm SFR_{UV}$ that is lower by $\sim 0.6$ dex.  In addition, the highest-$\rm M_*$ TKRS galaxies include objects with the highest SFRs in the sample.  This suggests that either outflow absorption strength is most closely correlated with SFR, or that there is evolution of outflows in $\rm \log M_* \sim 10$ galaxies between $z \sim 1.4$ and 1.  This will be discussed in greater detail in \S\ref{sec.discussion}. 

The coadd of early-type galaxies (as classified by G-$\rm M_{20}$) has very low $\rm S/N$ and is not useful for measuring outflow absorption.  $W_{diff}$ for the merger candidate and late-type subsamples are the same within the errors, at $W_{diff} \sim 0.3 - 0.5$ \AA; i.e., there does not appear to be a significant difference in the strength of outflow absorption in late-type galaxies and merger candidates (see Table 
\ref{tab.ew}).  
However, it is difficult to disentangle the effects of morphology and $\rm SFR_{UV}$ on outflow strength in this analysis.  
Studies of 
galaxy mergers (i.e., LIRGs and ULIRGs) in the local Universe show that they host exceptionally strong outflows \citep{Martin2005,Rupke2005b} in comparison to local late-type galaxies.  On the other hand, these merger remnants also have some of the highest SFRs at $z \sim 0$.  This degeneracy between morphology and SFR is broken at $z \sim 1$, where the majority of LIRGs have disk-like morphologies \citep{Melbourne2008}; thus galaxies at $z \sim 1$ may provide the ideal laboratory for investigating the effects of these two parameters on outflows.  Higher $\rm S/N$ spectra are required to examine these effects in greater detail.

 \begin{figure}
\includegraphics[angle=90,width=3.25in]{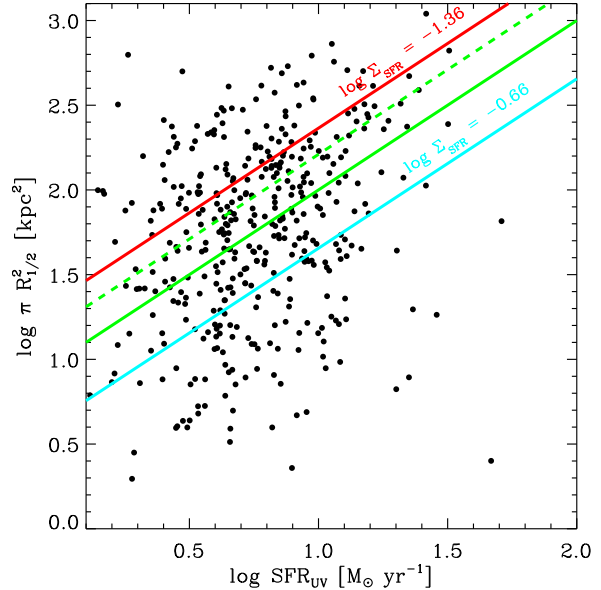}
\caption{$\rm \log \pi R_{1/2}^2$ vs. $\rm \log SFR_{UV}$ for the blue galaxies in the TKRS sample.  Lines show our subdivisions in $\rm \log \Sigma_{SFR}$.  All galaxies with $\rm \log \Sigma_{SFR} \le -1.36$ are above the red line; galaxies with $\rm-1.36 < \log \Sigma_{SFR} \le -0.66$ are between the red and cyan lines; and galaxies with $\rm \log \Sigma_{SFR} > -0.66$ are below the cyan line. A line of constant $\rm \log \Sigma_{SFR} = -1$ is shown in solid green.  The dashed green line indicates where galaxies in the sample lie with respect to a line of constant $\rm \log \Sigma_{SFR} = -1$ when the SFRs are increased by 0.21 dex.  
\label{fig.logrlogsfr}}
\end{figure}

\begin{figure}
\includegraphics[width=3.5in]{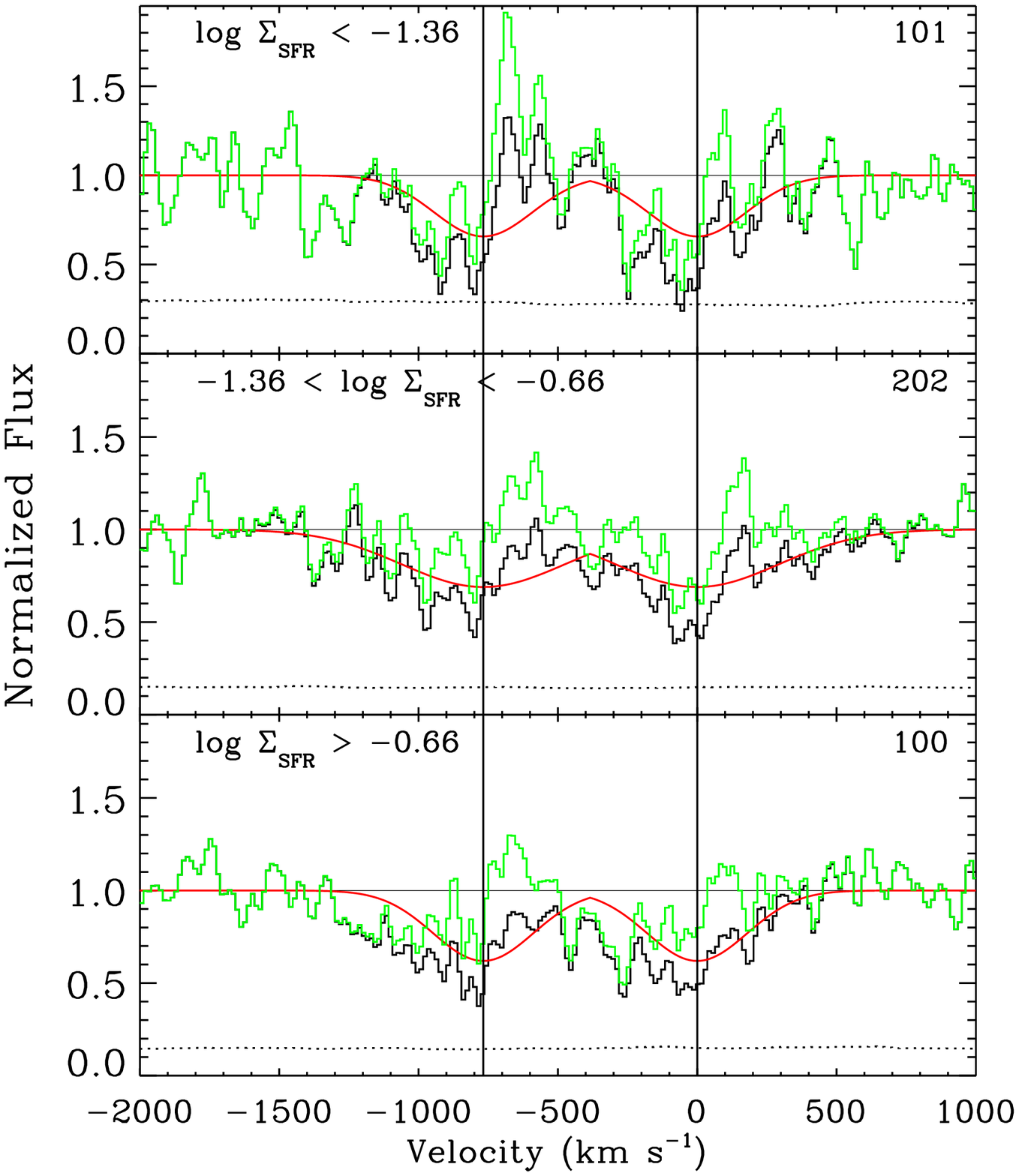}
\caption{Sections of the coadded spectra for subsamples divided by $\rm \Sigma_{SFR}$ around \ion{Mg}{2} (black).  The coadds have been normalized to the level of the continuum surrounding the absorption
lines as described in the text.  Colored lines are the same as in Figure~\ref{fig.decompsfrspec}.  The number of spectra in each subsample is shown in the upper right of each panel.\label{fig.decompsfrsdspec}}
\end{figure}

\subsection{Division by $\rm \Sigma_{SFR}$}\label{sec.sfrsd}
To test for a dependence of outflow strength on SFR surface density ($\rm \Sigma_{SFR}$) at $z > 0$, we assume that $\rm R_{SF,1/2} = R_{1/2}$, and combine this 
with our measure of $\rm SFR_{UV}$ to calculate the global (i.e., flux averaged) $\rm \Sigma_{SFR} = SFR/\pi R_{SF,1/2}^2$ for each object.   
Figure~\ref{fig.logrlogsfr} shows the distribution of $\rm \log SFR_{UV}$ and $\rm \log \pi R_{1/2}^2$ for the blue galaxies in the TKRS sample.  The lines show the 25th and 75th percentile values of $\rm \log \Sigma_{SFR}$ (-1.36, -0.66), which we use to subdivide the sample.   Figure~\ref{fig.decompsfrsdspec} shows the coadds of the spectra in these subsamples and the results of the decomposition procedure.     

\begin{figure}
\includegraphics[width=3.5in,angle=90]{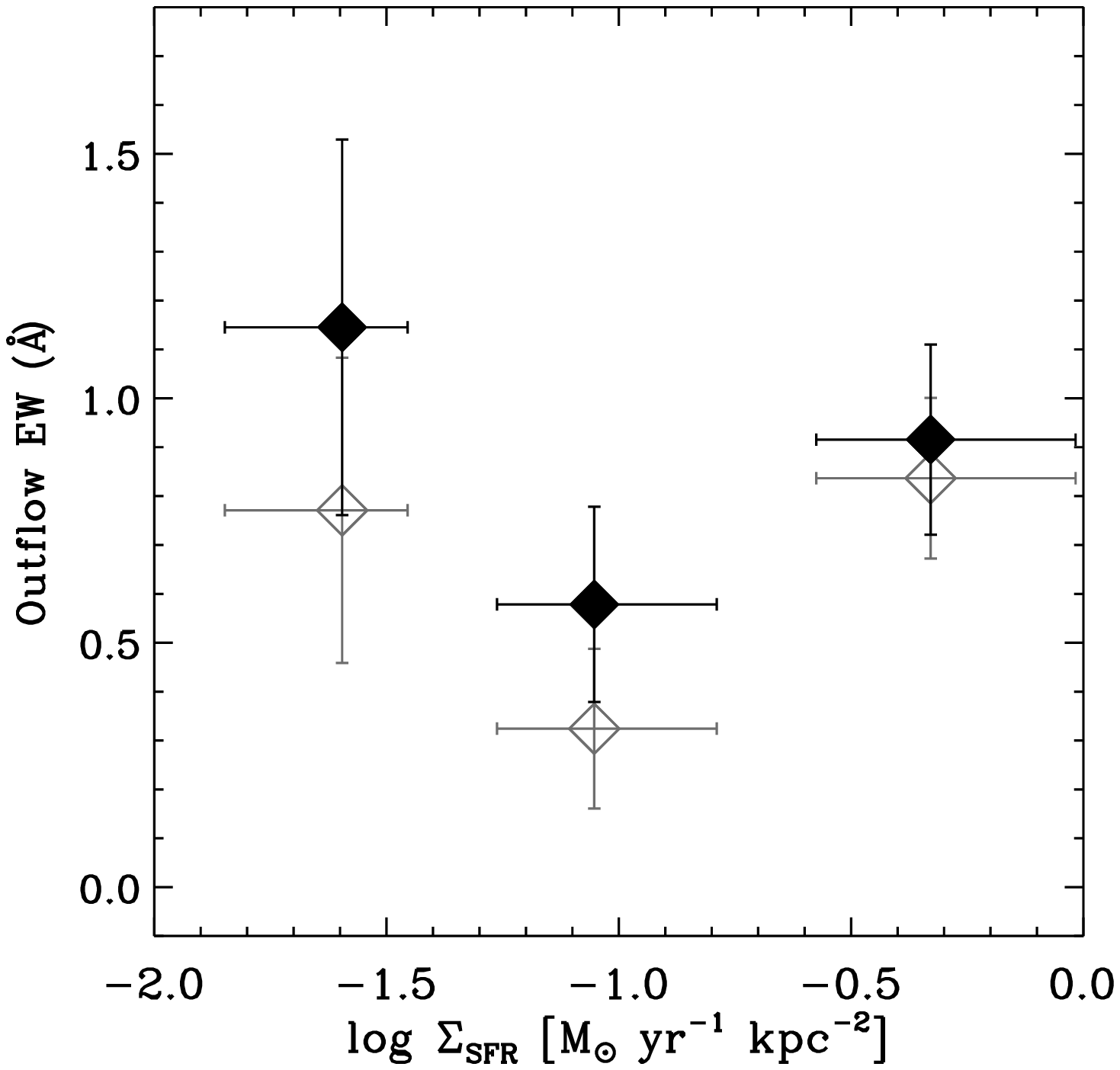}
\caption{ Outflow absorption strength vs.\ $\rm \Sigma_{SFR}$.  
Gray open points plot $W_{diff}$ measurements for the  
subsamples divided by the 25th and 75th percentile values of $\rm \Sigma_{SFR}$.  
Black points show the outflow EW results from the decomposition procedure for the same subsamples.  
\label{fig.bxcr_outflow_SFRSD}}

\end{figure}

Figure~\ref{fig.bxcr_outflow_SFRSD} shows $W_{diff}$ and outflow EW vs. $\rm \Sigma_{SFR}$ in each coadd. 
There is a slight increase in outflow absorption strength with $\rm \Sigma_{SFR}$ evident between the middle- and high-$\rm \Sigma_{SFR}$ subsamples.  The lowest-$\rm \Sigma_{SFR}$ subsample has $\rm S/N \sim 3~pix^{-1}$, and so here we are sensitive only to high velocity outflows with large velocity widths.  As discussed in \S\ref{sec.sn}, we can be confident that outflows are present in subsamples with low $\rm S/N$ coadds only when the measured outflow EW $\gtrsim 1.2-1.3$ \AA.  The outflow EW we measure for the lowest-$\rm \Sigma_{SFR}$ bin is only slightly below this limit, and is therefore only suggestive of the presence of outflow.  Comparing Figures~\ref{fig.bxcr_outflow_SFR_M} and \ref{fig.bxcr_outflow_SFRSD}, it appears that the outflow absorption strength is higher in the high-$\rm SFR_{UV}$ subsample than in the high-$\rm \Sigma_{SFR}$ subsample, suggesting that outflow absorption strength is more strongly correlated with absolute $\rm SFR_{UV}$ than with $\rm \Sigma_{SFR}$.

We also consider our results in the context of the suggested local ``threshold" for driving outflows, \thresh.  Because the existence of and a precise value for a strict threshold $\rm \Sigma_{SFR}$ for driving outflows have not yet been observationally established \citep[see, e.g.,][]{Strickland2004}, it is interesting to search below the suggested threshold for evidence of winds.  Given the uncertainty in our absolute SFR determinations, it is difficult to differentiate which of our individual galaxies have $\rm \Sigma_{SFR}$ above or below \thresheq.  The solid green line in Figure~\ref{fig.logrlogsfr} shows a line of constant \thresheq \ for comparison with the distribution of our sample in $\rm \pi R_{1/2}^2$-$\rm SFR_{UV}$ space.  From the placement of this line, it appears that many of the galaxies in the middle-$\rm \Sigma_{SFR}$ subsample have $\rm \Sigma_{SFR}$ values below the threshold.  If we shift this line by -0.21 dex in $\rm SFR_{UV}$ to reflect our possible systematic underestimate of SFR (see \S\ref{sec.sfr}) as indicated by the dashed green line, a smaller fraction of the sample falls below the threshold.  A 0.21 dex correction to the $\rm \Sigma_{SFR}$ values of the sample is likely too large, as this offset is derived from a comparison between IR- and UV-based SFRs for the IR-detected galaxies only.  However, we apply this correction and create a new subsample with ``corrected" $\rm \Sigma_{SFR} < 0.1~M_{\odot}~yr^{-1}~kpc^{-2}$.  The coadd of these spectra (referred to as the $\rm \log \Sigma_{SFR, Corrected}$ sample in Table~\ref{tab.ew}) yields $\rm S/N \sim 4.7~pix^{-1}$, $W_{diff} = 0.64 \pm 0.24$ \AA \ and outflow $\rm EW = 0.90 \pm 0.29$ \AA.  This implies that some galaxies below a \emph{flux averaged} \thresheq \ drive outflows.  Because there is a $1\sigma$ dispersion in the comparison between $\rm SFR_{UV}$ and $\rm SFR(IR)$ of 0.43 dex, some fraction of galaxies in this subsample do have true $\rm \Sigma_{SFR}$
values above the threshold; however, the absorption line profile in the coadded spectrum reflects the mean absorption properties of the subsample, which has $\rm \Sigma_{SFR}$ below the threshold in the mean.

\subsection{Redshift Dependence}
\begin{figure*}
\includegraphics[angle=90,width=7in,viewport=-10 0 410 730,clip]{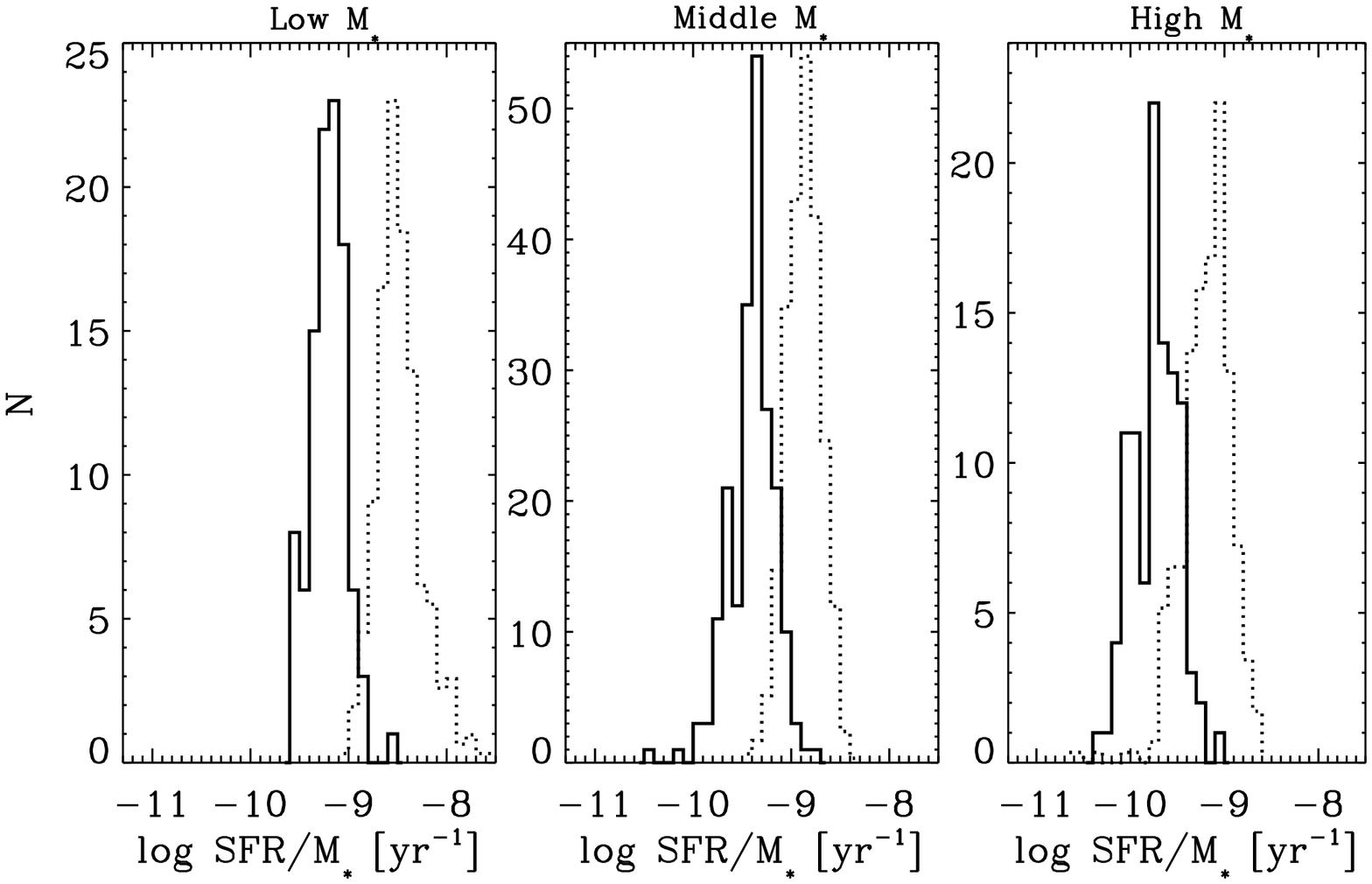}
\caption{Normalized SSFR distributions for galaxies in the low-$\rm M_{*}$ (left), middle-$\rm M_*$ (middle), and high-$\rm M_*$ (right) subsamples.  The solid histograms show the SSFR distributions for TKRS galaxies, and the dotted histograms show the SSFR distributions for the W09 $\rm M_*$-divided subsamples, normalized to have the same maximum values as the corresponding TKRS distributions. 
\label{fig.ssfrhist}}
\end{figure*}

As noted in \S\ref{sec.sfr}, the TKRS and W09 samples span a similar range in $\rm M_*$, but the TKRS sample is offset to lower SFRs by $\sim 0.6$ dex.  
To examine evolution in outflow properties with redshift, we compare our outflow measurements to those from W09.   
The 25th and 75th-percentile $\rm M_*$ divisions of our sample are quite close to the divisions used in W09
($\rm \log M_* < 9.88, 9.88 < \log M_* < 10.45$, and $\rm \log M_* > 10.45$).  To further investigate similarities between the two sets of subsamples, we compare the distributions of specific SFR ($\rm SFR/M_*$; SSFR).  Figure~\ref{fig.ssfrhist} shows the 
SSFR distribution of TKRS (solid line) and DEEP2 (dotted line) galaxies in each bin.  Each subsample has a symmetric distribution in SSFR at both redshifts; these distributions are simply offset from each other because of the decline in global SFR with decreasing redshift.  The symmetry of the distributions suggests that none of the subsamples are severely contaminated with non-star-forming galaxies.  
See \S\ref{sec.taus} and Figure~\ref{fig.taus} for further comparison of the SSFR-$\rm M_*$ relations of the W09 and TKRS samples and a discussion of the evolutionary connection between the galaxies at the two redshifts.

\begin{figure}

\includegraphics[angle=90,width=3.25in]{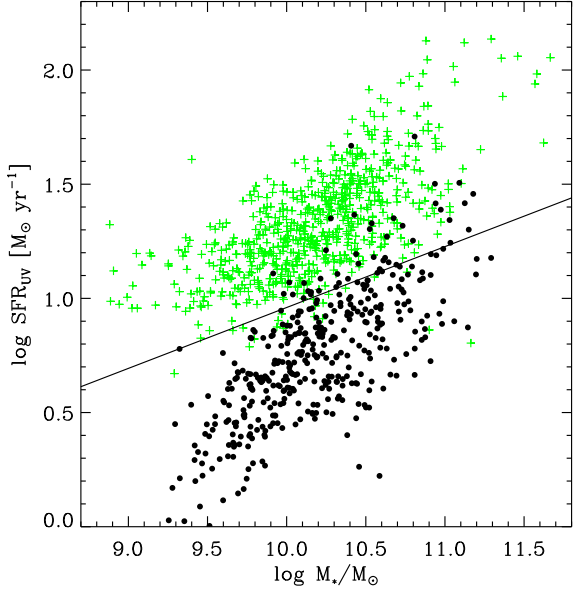}
\caption{$\rm \log SFR_{UV}$ vs.\ $\rm \log M_{*}$ for the full TKRS sample (in black) and a randomly selected portion of the DEEP2 sample (in green) from W09.  The line shows the cut applied to the TKRS sample for the purposes of selecting TKRS galaxies with $\rm SFR_{UV}$s similar to the $\rm SFR_{UV}$ values in the W09 sample.  \label{fig.d2masscut}}
\end{figure}

\begin{figure}

\includegraphics[width=2.85in,angle=90]{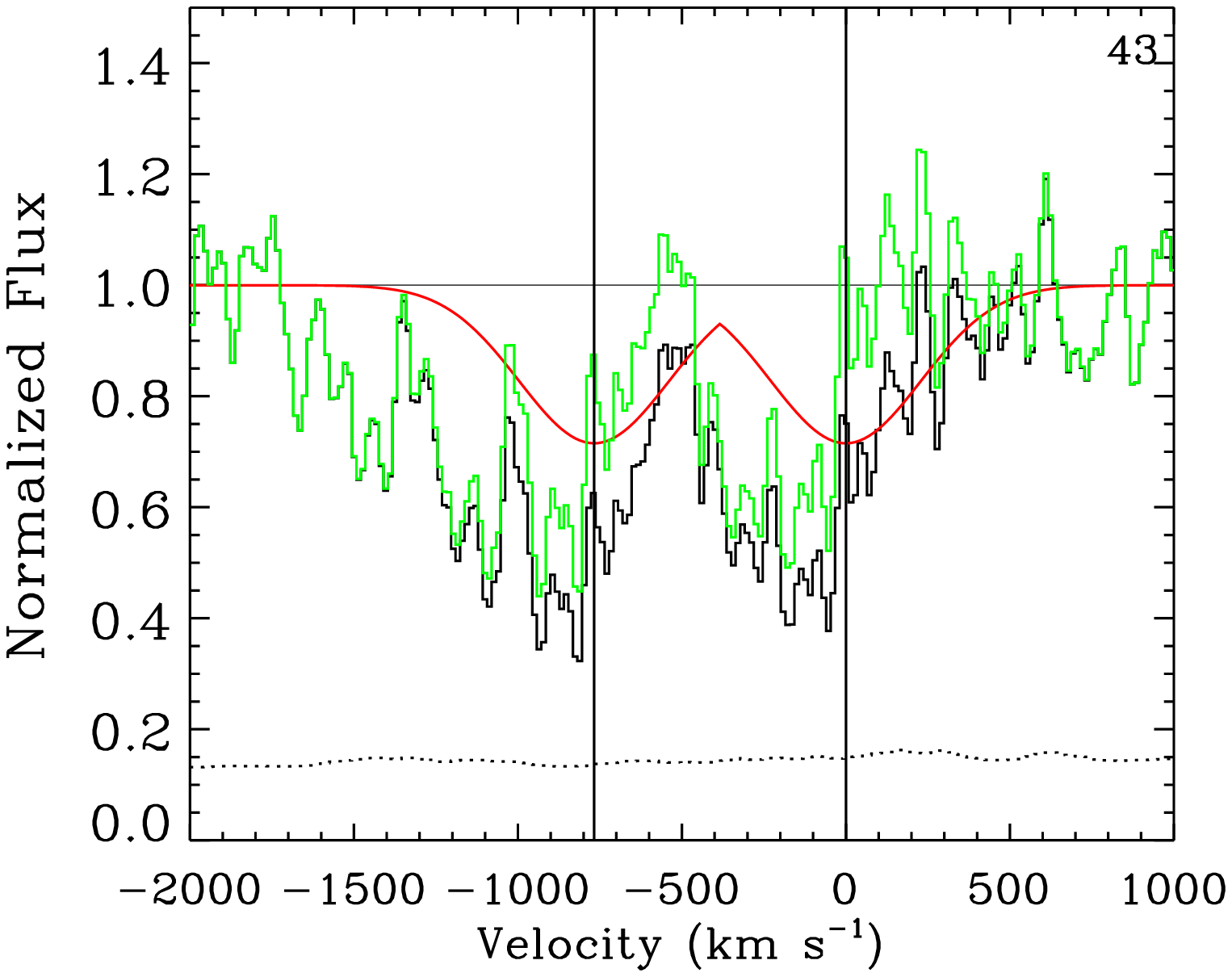}
\caption{Sections of the high-SFR ``DEEP2-like" sample coadd around \ion{Mg}{2}.  The coadd
has been normalized to the level of the continuum surrounding the absorption
lines as described in the text.   The colored lines are the same as in Figure~\ref{fig.decompsfrspec}.  The number of spectra in the sample is shown in the upper right.  \label{fig.decompd2sfrcompspec}}
\end{figure}

Although the TKRS and W09 samples are nearly disjoint in $\rm SFR_{UV}$-$\rm M_*$ space, we may construct a comparison subsample of 43 TKRS galaxies which lie above the lower envelope of the SFRs for the W09 galaxies.
Figure~\ref{fig.d2masscut} shows the TKRS (black circles) sample and a random selection of half of the W09 (green crosses) sample in $\rm \log SFR_{UV}$-$\rm \log M_*$ space.  We include all TKRS galaxies above the solid line in a high-$\rm SFR_{UV}$, ``DEEP2-like" sample.
Figure~\ref{fig.decompd2sfrcompspec} shows the coadd and decomposition analysis of this subsample.

The red point in Figure~\ref{fig.bxcr_outflow_SFR_M} shows results for this DEEP2-like sample. 
This sample as well as the high-$\rm SFR_{UV}$ TKRS subsample have outflow EWs similar to or larger than the W09 coadds.  This confirms our previous finding that galaxies at $z \sim 1.4$ and 1 with similar SFRs have strong outflow absorption, and that galaxies with lower SFRs have weaker outflows.  We note that the galaxies in the highest-$\rm M_*$ TKRS subsample have a median $\rm \log SFR_{UV} = 0.99$, with 68\% of the subsample in the range $\rm 0.78 < \log SFR_{UV} < 1.22$.  These galaxies therefore have $\rm SFR_{UV}$ values in a range similar to the W09 sample; the high outflow EW for this subsample is consistent with the measurements for our other high-$\rm SFR_{UV}$ subsamples.

 \section{\ion{Fe}{2} 2586, 2600 Absorption}\label{sec.feiimain}

 We now focus on measurements of \ion{Fe}{2} absorption in our spectra.  
A constraint on the extent to which \ion{Fe}{2} 2586, 2600 absorption is detected with \ion{Mg}{2} in outflowing gas is valuable for a number of reasons, as discussed in \citet{MartinBouche2009}.  While \ion{Mg}{2} is found in a wide range of gas densities, \ion{Fe}{2} is present only at higher densities (in more neutral gas), and so may provide information about outflow gas density.  Second, the \ion{Fe}{2} 2586, 2600 transitions have oscillator strengths 0.11 and 0.39 times that of \ion{Mg}{2} 2796, and Fe is less abundant in general than Mg; thus \ion{Fe}{2} absorption lines are generally less strongly saturated than \ion{Mg}{2} lines.   They may therefore be used to place more stringent constraints on the cool outflow column density.  For instance, \citet{MartinBouche2009} find that the \ion{Fe}{2} absorption in ULIRG outflows at $z \sim 0.25$ requires optical depths 2-3 times larger than those derived from analysis of \ion{Mg}{2}.  Finally, because Mg is generated in Type II supernovae in younger stellar populations than Fe, comparison of the relative abundances of \ion{Mg}{2} and \ion{Fe}{2} in the outflow may constrain the enrichment history of the gas.

The \ion{Fe}{2} 2600 transition is always weaker than \ion{Mg}{2} 2796 in QSO absorption line systems, with at most $2/3$ the EW of \ion{Mg}{2} 2796 \citep{Churchill2000}.  This is also true in stellar atmospheres.  \ion{Fe}{2} 2600 reaches its maximum absorption strength in F9 - G2 stars, while \ion{Mg}{2} reaches its maximum strength in F8 - G1 stars \citep[][although \emph{asymmetric} \ion{Mg}{2} absorption is found in B-F stars as mentioned in \S\ref{sec.analysis}]{Kinney1993}.  In early type stars, \ion{Fe}{2} lines do not exhibit P-Cygni profiles or asymmetric absorption; instead if mass-loss effects are observed the lines are simply blueshifted \citep{Snow1994}.  

We use UVBLUE theoretical stellar spectra \citep{Rodriguez-Merino2005} to measure the relative EWs of these transitions in stellar atmospheres.  We first normalize these spectra, calculating the continuum in the same regions used for our galaxy coadds.  We measure EWs between $\rm -500~km~s^{-1} < v < 500~km~s^{-1}$ for \ion{Fe}{2} 2600 and between $\rm -500~km~s^{-1} < v < 385~km~s^{-1}$ for \ion{Mg}{2} 2796.  Note that these spectral regions contain absorption not only from the named transitions, but from several weaker transitions also found in stellar atmospheres (the \ion{Fe}{2}* 2599 transition, for example).  Using solar metallicity models, we obtain $\rm EW (2796) \sim 6.8$ \AA \ and $\rm EW (2600) \sim 3.1$ \AA \ for a solar type star, yielding a ratio of $\sim 2.2$.  As the atmospheres become hotter, the $\rm EWs$ of both ions decrease, although the $\rm EW (2600)$ value decreases much faster with increasing temperature than $\rm EW (2796)$.

We now measure \ion{Fe}{2} EWs and analyze \ion{Fe}{2} line profiles to constrain the origin of the \ion{Fe}{2} absorption in our spectra.  

\subsection{\ion{Fe}{2} Absorption in TKRS2158}
As noted in \S\ref{sec.TKRS2158}, we observe blueshifted \ion{Fe}{2} lines in the spectrum of TKRS2158.  Velocity offsets are given in Table~\ref{tab.veloff2158}.  Velocities are slightly inconsistent among the various \ion{Fe}{2} transitions; this may be due to poorly subtracted sky emission, intrinsic \ion{Fe}{2}* emission, or other sources of noise in the spectrum.  In general, \ion{Fe}{2} velocities are slightly lower than the offsets measured for \ion{Mg}{2}.  This may occur if \ion{Fe}{2} absorption is weaker at the highest velocities, and / or if \ion{Mg}{2} emission fills in the line profile near systemic velocity and effectively shifts the line center further to the blue.  From Figure~\ref{fig.2158uvlinesstamps} we see that a few of the \ion{Fe}{2} lines extend to the same depth as the \ion{Mg}{2} lines.  We conclude that \ion{Fe}{2} traces outflowing gas in this object, although we cannot confirm that it traces gas in the same velocity range as \ion{Mg}{2}. 

\subsection{\ion{Fe}{2} Absorption in Coadded Spectra}\label{sec.feii}
To analyze \ion{Fe}{2} absorption in our galaxy spectra, we create a new subsample, selecting objects with $ z \gtrsim 0.8$ such that spectral coverage of \ion{Fe}{2} $\lambda \lambda 2586, 2600$ is available, and further selecting objects with $\rm \log SFR_{UV} \geq 1$, for a final subsample size of 66 objects.  This latter cut selects objects with the bluest colors and the strongest continua near 2600 \AA, which minimizes unphysical effects due to poorly determined sky levels.  Coadding spectra of fainter galaxies in the sample yields \ion{Fe}{2} absorption with unphysical line profiles, and thus are excluded from this analysis.

\begin{figure}

\includegraphics[width=3.25in,angle=90]{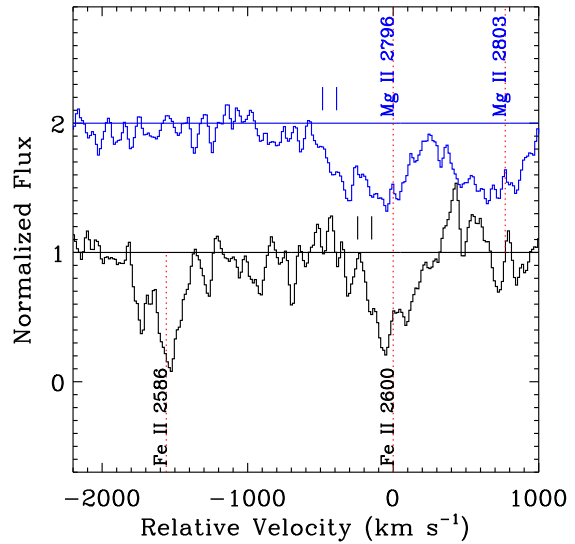}
\caption{Sections of the coadd of galaxy spectra used in the \ion{Fe}{2} analysis with $\rm \log SFR_{UV} \ge 1$.  The black spectrum shows the region surrounding \ion{Fe}{2} 2586, 2600, and the blue spectrum shows the region surrounding \ion{Mg}{2} 2796, 2803.  Coadds have been normalized to the level of the continuum surrounding the absorption
transitions, indicated with the horizontal lines.  The section surrounding \ion{Mg}{2} has been offset by +1 in flux.  The velocity for the black spectrum is measured relative to the rest velocity of the $2600~\rm \AA$ line; for the blue spectrum it is measured relative to the rest velocity of the $2796~\rm \AA$ line.  Vertical red dotted lines show rest velocities of all lines of interest.  Vertical blue and black marks indicate the velocities at which the absorption decreases to 80\% and 60\% of the continuum level for the \ion{Mg}{2} 2796 and \ion{Fe}{2} 2600 lines, respectively.\label{fig.op_allspec}}
\end{figure}

Figure~\ref{fig.op_allspec} shows \ion{Fe}{2} and \ion{Mg}{2} line profiles for this subsample.  
We note that the \ion{Fe}{2} lines are slightly deeper than the \ion{Mg}{2} lines in the coadd, in contrast to examples from QSO absorption line systems \citep{Churchill2000} and stellar atmospheres.   
This inconsistency suggests that a) the depth of \ion{Fe}{2} is affected by unphysical artifacts resulting from poor sky subtraction even in this subsample of the brightest galaxies, b) \ion{Fe}{2}-absorbing gas has a larger $C_f$ than \ion{Mg}{2}, or c) the \ion{Mg}{2} absorption is being filled in by emission in the same transition.  

We measure the EW in the \ion{Fe}{2} lines in the coadded spectrum between $\rm -300~ km~s^{-1} < v < 300~ km~s^{-1}$, as well as on the blue ($\rm -300~ km~s^{-1} < v < 0~ km~s^{-1}$) and red ($\rm 0~km~s^{-1} < v < 300~km~s^{-1}$) sides of each.
The total EWs for each line are approximately the same ($\rm EW(2586) = 1.99 \pm 0.15$ \AA \ and $\rm EW(2600) = 1.86 \pm 0.15$ \AA), indicating that \ion{Fe}{2} is completely saturated in this coadd.
The total EWs of the \ion{Mg}{2} absorption lines in the same velocity range are significantly larger ($\rm EW(2796) = 2.36 \pm 0.10$ \AA), yielding $\rm EW(2796) / EW(2600) \sim 1.2$, a slightly smaller ratio than expected from stellar atmospheres and QSO absorption lines.  This again may be due to emission filling in the \ion{Mg}{2} absorption line profiles.  We measure a larger EW on the blue side of the 2600 \AA \ line than on the red side ($\rm EW(2600, blue) = 1.12 \pm 0.11$ \AA \ and $\rm EW(2600, red) = 0.79 \pm 0.10$ \AA).  This in consistent with a scenario in which the \ion{Fe}{2} absorbing gas is outflowing.  We find that the EW on the blue and red sides of the 2586 \AA \ line in the coadd are consistent within the errors ($\rm EW(2586, blue) = 0.99 \pm 0.11$ \AA \ and $\rm EW(2586, red) = 1.10 \pm 0.11$ \AA).  We cannot conclude that \ion{Fe}{2} absorbing gas is outflowing from analysis of this line alone; however, the symmetry of the profile does not rule out an outflow scenario.  The oscillator strength of this transition may simply be too low for it to obviously trace blueshifted gas.  In contrast, \citet{MartinBouche2009} find that both the \ion{Fe}{2} 2586 and 2600 \AA \ lines are saturated and have similar line profiles in ULIRG outflows.  See \S\ref{sec.feiicolumn} for a discussion of the upper limit on $N$(H) in the outflow derived from \ion{Fe}{2} EWs.

We also examine the relative velocity extent of the \ion{Fe}{2} and \ion{Mg}{2} absorption lines.  In Figure~\ref{fig.op_allspec}, 
the \ion{Mg}{2} 2796 absorption extends to larger velocities than the \ion{Fe}{2} 2600 absorption.  The \ion{Mg}{2} 2796 line has a high velocity tail which decreases in strength gradually with increasing velocity offset, whereas the \ion{Fe}{2} 2600 line has no ``tail" to high velocities.  If the \ion{Fe}{2} absorbing gas is indeed outflowing, this suggests that the majority of this gas does not attain velocities as high as the \ion{Mg}{2} absorbing gas.

To quantify this, we measure the relative velocities where the profiles of each of these lines reaches a threshold amount of absorption.  We first smooth the coadds, and calculate the velocity at which the absorption decreases to 80\% and 60\% of the continuum level.  These velocities are marked in Figure~\ref{fig.op_allspec}, and are closer to the center of the line in the case of \ion{Fe}{2}, confirming our finding that \ion{Mg}{2} absorption extends to higher velocities.  If we make the assumption that Fe and Mg have the same relative abundances and $C_f$ at all gas velocities, this suggests that the density of the outflowing clouds decreases with increasing velocity, as the \ion{Fe}{2} column decreases with density \citep[see, e.g.,][for ionization modeling of \ion{Mg}{2} absorbers]{Narayanan2008}.  

\section{Discussion}\label{sec.discussion}

\subsection{Physical Characteristics of Outflows}\label{sec.column}

 \subsubsection{\ion{Mg}{2}}
 Similarly to W09, 
 we use the ratio of the EWs on the blue sides of the \ion{Mg}{2} lines to estimate a lower limit on the column density ($N(\mathrm{Mg~II})$) in the outflowing gas.  The oscillator strengths of the lines in the \ion{Mg}{2} doublet are in the ratio $2:1$, and in the optically thin case the ratio of the EWs of the lines will also be $2:1$.  As the optical depth at line center ($\tau_0$) increases and the lines become saturated, the EW ratio approaches $1:1$; therefore the EW ratio can be used to constrain $\tau_0$.  Once $\tau_0$ is known, $N(\mathrm{Mg~II})$ can be calculated using the equation \citep{Spitzer1968}
\begin{eqnarray}
	\log N &=& \log \frac{EW}{\lambda} - \log \frac{2 F(\tau_0)} {\pi^{1/2} \tau_0} \nonumber \\ 
	& & { } - \log \lambda f - \log C_f + 20.053,
\end{eqnarray}
 adjusted to include the effect of the covering fraction, $C_f$, where $\lambda$ and $\rm EW$ are in \AA, and $N$ is in atoms  $\rm cm ^{-2}$.  $F(\tau_0)$ is given by:
 \begin{eqnarray}
 	F(\tau_0) = \int_0^{+\infty} (1 - e ^{-\tau _0 \exp(-x^2)})~ dx.
\end{eqnarray}
The EW ratio in the \ion{Mg}{2} doublet is almost exactly equivalent to the ``doublet ratio",
$F(2 \tau_0) / F(\tau_0)$.  After this doublet ratio is calculated by taking the ratio of the EWs, one may numerically solve for $\tau_0$.  This method is strictly appropriate only when one absorbing cloud is considered.  However, it has been found to yield good results even when the absorption is caused by a number of clouds if the optical depth in the weaker line is $\tau_0 < 5$ \citep{Jenkins1986}.  

We measure the EWs for the coadds in the interval $\rm -384~ km~s^{-1} < v < -200~ km~s^{-1}$ for each line in the doublet; these measurements are listed in Table~\ref{tab.ew}.  We choose this interval to avoid measuring absorption at the systemic velocity, which is uncertain to within $200 \mkms$ (see \S\ref{sec.stacking}), and to avoid including absorption from the red side of the 2796 \AA \ line in our 2803 \AA \ line measurement.  For the coadd of the entire sample, the EW ratio is $1.09 \pm 0.16$.   
This yields 
$\tau_{0, 2803} \sim 3 - \infty$, where the lower limit corresponds to the $1\sigma$ upper limit on the EW ratio.  In the case of the high-$\rm M_*$ coadd, the doublet ratio is $1.28 \pm 0.16$, which results in 
$\tau_{0, 2803} \sim 1.4 - 10$.  Using the equation for $N$ above, and assuming $C_f = 1$, we find $\log N(\rm Mg~II) > 13.7$ for both the full sample and the high-$\rm M_*$ subsample.   
Assuming $C_f = 0.5$ increases $\log N$ by 0.3 dex. 

 To estimate a lower limit to $N(\mathrm{H})$ in the outflow, we assume a more conservative value of $C_f = 1$, and we assume $N(\mathrm{Mg}) = N(\mathrm{Mg~II})$; i.e., we do not apply an ionization correction.  We assume the solar value for the abundance of Mg, $\log \rm Mg / H = -4.42$, and a factor of -1.2 dex Mg depletion onto dust, measured in the local ISM \citep{SavageSembach1996}.  This results in the estimate $\log N(\rm H) \gtrsim 19.3$.  We emphasize that this is a very conservative lower limit on the column of outflowing gas because of our assumption about $C_f$, our neglect of ionization corrections, and because our method underestimates $N$ for highly saturated \ion{Mg}{2}-absorbing velocity components.  In addition, the effect of emission from the \ion{Mg}{2} 2796 transition is to reduce the EW measured blueward of the \ion{Mg}{2} 2803 transition, further reducing the calculated $\log N(\rm H)$.  These results are nearly an order of magnitude smaller than the outflow column obtained in W09 ($\log N(\rm H) = 20.1$); however, this latter measurement was calculated using the EW ratio in the outflow line profile rather than in the observed profile, and used a wider range of velocities ($\rm -768~\mkms < v < 0~\mkms$).  This yielded higher EWs in each line and generally higher optical depths than our $1\sigma$ lower limits on $\tau$.
 
 We estimate the mass outflow rate by assuming a specific geometry for the outflowing gas.  For a thin shell, the mass outflow rate is given by 
 \begin{eqnarray}
	\dot M \approx 22 ~\mathrm{M_{\odot} yr^{-1}} C_f \frac{N(\mathrm{H})}{10^{20}~ \mathrm{cm^{-2}}} \frac{R}{\mathrm{5~ kpc}} \frac{v}{300~ \mathrm{km~ s^{-1}}}
\end{eqnarray} 
from W09.  The assumption for $C_f$ is unimportant here, so long as it matches what was assumed for the calculation of $N(\rm H)$  (i.e., the factor of $C_f$ cancels out).   We have no constraint on the radial extent of the wind from our data ($R$), except that it is likely comparable to the size of the galaxies, because $C_f$ is high.    
For these purposes, we assume a minimum radius for the shell equal to the median half-light radius for the galaxy sample, 4.1 kpc.  
We use the velocity at 80\% of the continuum in the coadd of all galaxies calculated as described in \S\ref{sec.feii}, $-334~\rm km~s^{-1}$.    
Note that this is \emph{not} the same velocity measurement used in W09, who measured the velocity at 50\% opacity in the outflow component after decomposing the line profile.
The resulting mass outflow rate is $\dot M \gtrsim 5~ \rm M_{\odot}~yr^{-1}$.  As in previous work \citep[e.g.,][]{Martin1999,Rupke2005b,Martin2005}, we find an $\dot M$ on the same order as the SFR of the sample ($\sim 1 - 30~\rm M_{\odot}~yr^{-1}$).  We note that $\dot M$ in this sample is lower than in the W09 sample at $z \sim 1.4$, and that this is due to lower outflow $N(\rm H)$ at the lower redshifts.  This suggests that the galaxies at lower $z$ are less effective in driving outflows than at $z \sim 1.4$.  However, we reiterate that the outflow $N(\rm H)$ in both this work and in W09 may be substantially underestimated.

 \subsubsection{\ion{Fe}{2}}\label{sec.feiicolumn}
Because the profiles of the two \ion{Fe}{2} lines do not have an asymmetric blue wing, we hypothesize that either these lines are too weak to trace outflows at velocities $ \gtrsim 300~\rm km~s^{-1}$, or alternatively that \ion{Fe}{2} simply is not present in outflowing gas at these velocities.  However, assuming the absorption on the blue sides of these lines at \emph{lower} relative velocities is due to outflowing gas, we again use an EW ratio, this time in the velocity range $-300~\rm km~s^{-1} < v < -200~km~s^{-1}$, to calculate a lower limit on the outflow column in the \ion{Fe}{2} transition.  
The EWs of the two lines in this velocity range for the coadd we use to analyze \ion{Fe}{2} are both $\rm EW = 0.14 \pm 0.06$, yielding an observed EW ratio of $1.0 \pm 0.7$.   
The two \ion{Fe}{2} lines have oscillator strengths in the ratio 3.475:1.    
We numerically solve the equation $F(3.475 \tau_0) / F(\tau_0) = 1.0 \pm 0.7$ for $\tau_{0,2586}$ and obtain the limit $\tau_{0,2586} \gtrsim 1.5$.  
Using $\log \lambda_{2586} f = 2.252$ \citep{Morton2003} and $\rm EW_{blue} (2586) = 0.08 - 0.2$ \AA, we find $\log N(\rm Fe~ II) \gtrsim 13.5 - 13.9$.  Assuming solar abundance, no ionization correction and a factor of $1/10$ depletion onto dust, we obtain $\log N(\mathrm{H}) \sim 19.0 - 19.4$, in good agreement with our result from \ion{Mg}{2}.  In this case, analysis of the \ion{Fe}{2} profiles does not yield a more stringent constraint on the outflow column than analysis of \ion{Mg}{2}.   

We may compare column densities measured in different relative velocity ranges for these lines if we assume that the distribution of $N$ in individual clouds changes smoothly with changing velocity \emph{and} if we assume that the $C_f$ of the outflow is constant at all velocities.  We calculate a $3\sigma$ upper limit of $\sim 0.3$ \AA \ on the EW of absorption at velocities $\rm -500~km~s^{-1} < v < -300~km~s^{-1}$ with respect to both \ion{Fe}{2} absorption lines.  We use this measurement to set an upper limit on the outflow column at these velocities by assuming that this gas must have an optical depth lower than the limit determined above for the range $\rm -300~km~s^{-1} < v < -200~km~s^{-1}$.  Again we use the equation for $\log N$ given in the previous section and apply it to the case of the 2600 \AA \ line, with $\tau_{0,2600} = 3.475 \tau_{0,2586} = 5.2$ and $\log \lambda_{2600} f = 2.793$ \citep{Morton2003}.  We find $\log N(\mathrm{Fe~II}) < 13.8$ and $\log N(\mathrm{H}) < 19.3$, using the same assumptions for abundance, ionization correction and dust depletion as above.  Note that this result requires that $C_f = 1$ at all velocities; this assumption does not hold for the ULIRG outflows discussed in \citet{MartinBouche2009}.

\subsection{Outflow Absorption Dependence on SFR and $\rm M_*$}\label{sec.disc_sfr_m}
We have shown that outflow absorption strength increases with $\rm SFR_{UV}$ and $\rm M_*$ in the TKRS sample.  This statement addresses the behavior of outflow absorption in the middle- and high-$\rm SFR_{UV}$ and $\rm M_*$ subsamples only (with $\rm S/N > 6~pix^{-1}$); in the subsamples with the lowest values of these parameters, the coadds have $\rm S/N \sim 3~pix^{-1}$ and only poorly constrain outflow properties.  It does appear that the lowest-$\rm M_*$ subsample has a large enough outflow EW to be significant ($> 1$ \AA; see discussion in \S\ref{sec.sn}); however, the low measurement of $W_{diff}$ is consistent with no outflow.

Interestingly, both the middle-$\rm SFR_{UV}$ and $\rm M_*$ subsamples have $W_{diff}$ and outflow EW measurements $< 0.4$ \AA, consistent with no outflow.  These coadds have $\rm S/N \sim 6~pix^{-1}$; in the simulated spectra discussed in \S\ref{sec.sn}, saturated (``strong") outflows with $v \gtrsim 100 \mkms$ and $b_D \gtrsim 150 \mkms$ yield values of outflow EW greater than 0.85 \AA \ with 95\% confidence.  Certainly weaker outflows with smaller $b_D$, velocities $< 100 \mkms$, and with column densities such that the total $\rm EW($\ion{Mg}{2} 2796$) \lesssim 3.5$ \AA, would not be detected in these coadds, and such outflows could exist in these galaxies.  However, we find that the middle subsamples of galaxies with $\rm 0.555 < \log SFR_{UV} < 0.942$ and $\rm 9.86 < \log M_* < 10.49$ at $z \sim 1$ are not driving strong outflows.   

\begin{figure*}

\includegraphics[angle=90,width=7in]{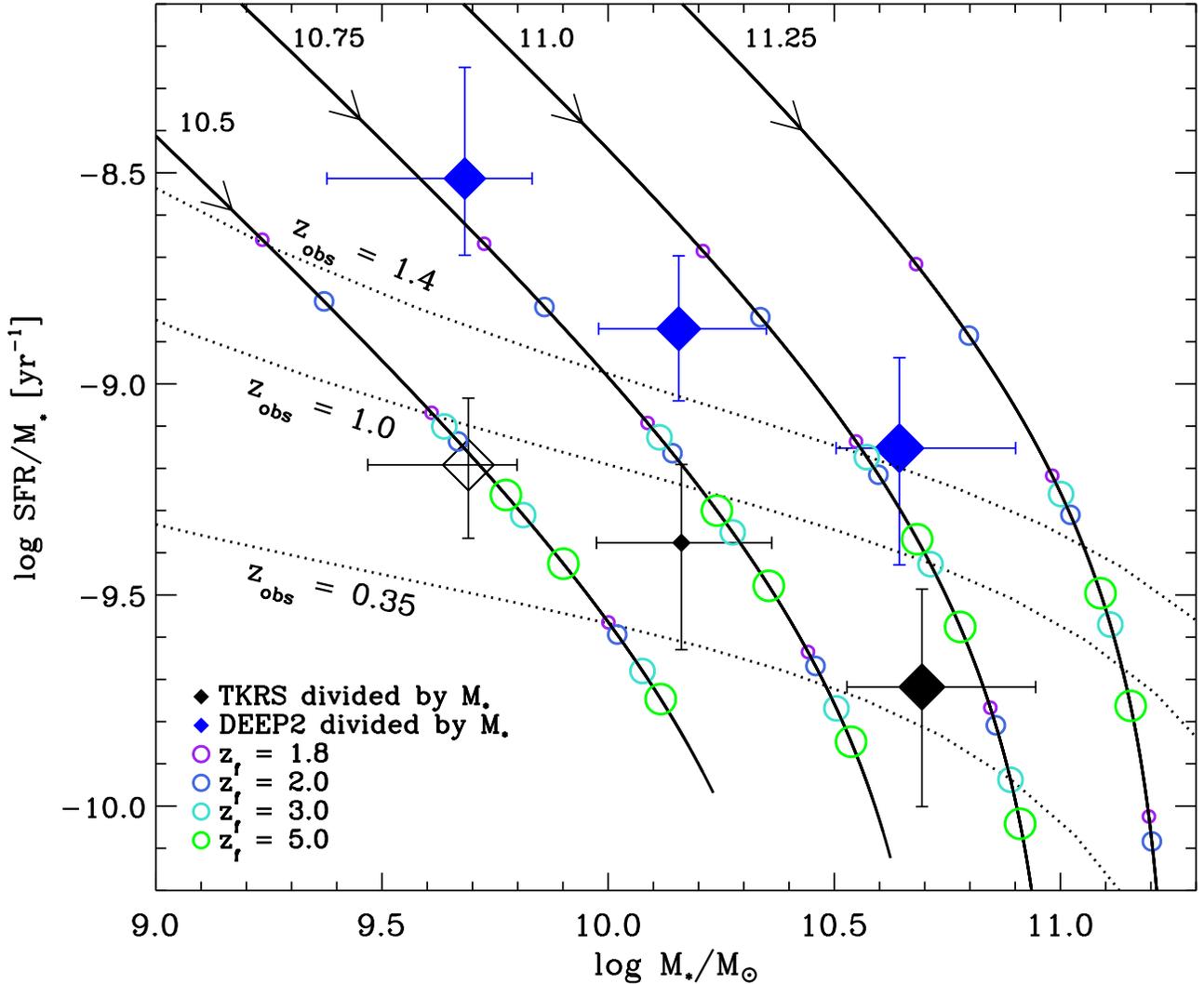}
\caption{Diamonds show SSFR vs.\ $\rm M_*$ for the W09 and TKRS subsamples divided by $\rm M_*$ (blue and black, respectively).  The point size is scaled to the square root of the outflow EW for each subsample.  Error bars show the $\pm 34$th-percentile values of SSFR and $\rm M_*$ for each subsample.  The open diamond indicates a subsample for which analysis of the low-$\rm S/N$ coadd provides only an upper limit on the outflow EW.  Dotted lines show fits to the SSFR-$\rm M_*$ main sequence derived in \citet{Noeske2007a} at $z_{obs}=0.35$, 1 and 1.4 (bottom to top).  Solid lines show the paths of $\tau$ model galaxies through the diagram with time for $\log M_b = 10.5$, 10.75, 11.0 and 11.25 (left to right).  The open circles along these lines indicate the location of individual models with different formation redshifts ($z_f$) at $z_{obs}=0.35,1.0$ and 1.4.  Each color corresponds to a model with a different formation redshift, and thus generally appears 3 times along each solid line (i.e., there is one circle of each color for each observed redshift).    
See text for more details.}\label{fig.taus}
\end{figure*}

Galaxies with $\rm M_*$ and $\rm SFR_{UV}$ above these ranges show clear evidence for outflow.  This includes galaxies in our subsample specifically selected to have similar $\rm SFR_{UV}$ and $\rm M_*$ to the W09 sample at $z \sim 1.4$ (the ``DEEP2-like" subsample).  It also includes the galaxies in the W09 sample, which all have $\rm \log SFR_{UV} \gtrsim 1$.  Our middle-$\rm M_*$ subsample overlaps with the W09 sample in $\rm M_*$ but falls below it in $\rm SFR_{UV}$, suggesting that outflow absorption strength is more closely associated with SFR than with $\rm M_*$ at $z = 1-1.4$.  We also note that outflow absorption strength in general does not increase with increasing specific SFR (SSFR).  Figure~\ref{fig.taus} shows the SSFR-$\rm M_*$ distribution of the TKRS and W09 $\rm M_*$-divided samples with black and blue diamonds, respectively.  Although the middle-$\rm M_*$ TKRS subsample has higher SSFR than the high-$\rm M_*$ TKRS subsample, it exhibits no outflow signature (see Figures~\ref{fig.decompsfrspec} and \ref{fig.bxcr_outflow_SFR_M}).

It is interesting to consider whether outflowing gas will escape from these high-$\rm M_*$ and high-$\rm SFR_{UV}$ galaxies.  The escape velocity of galaxies can be estimated from the width of the [\ion{O}{2}] emission line as in W09.  The [\ion{O}{2}] linewidths ($\sigma$) of TKRS galaxies were measured in \citet{Weiner2006}.  
We find that the median $\sigma$ for both our high-$\rm M_*$ and high-$\rm SFR_{UV}$ subsamples is $\sigma (\mathrm{[O~II]}) = 80~\rm km~s^{-1}$, and use the relation from W09:
\begin{eqnarray}
	V_{escape} \approx 5-6\sigma (\rm [O~II]).
\end{eqnarray}
This yields an escape velocity $\sim 480~\rm km~s^{-1}$.  From the outflow components plotted in green in the bottom panels of Figure~\ref{fig.decompsfrspec}, we see that in the high-$\rm SFR_{UV}$ subsample, which is most like the W09 sample in terms of SFR, there appears to be a tail of \ion{Mg}{2} absorption beyond the escape velocity.  In the other subsamples, there is no compelling evidence of gas beyond the escape velocity.  
We conclude that is it unlikely that the gas detected in absorption is able to escape from most of the galaxies in our sample.    
This differs from the results of W09, who find that the absorption line profile extends well beyond the median escape velocity for each of their $\rm M_*$-divided subsamples.  The profile extends to particularly high velocities in the highest-$\rm M_*$ galaxies; the velocity where the profile is at 90\% of the continuum level exceeds the escape velocities for $> 84$\% of the subsample with $\rm \log M_* > 10.45$.  We cannot conclude from this analysis, however, that cool \ion{Mg}{2}- absorbing gas does not escape from the bulk of the TKRS sample at $z \sim 1$.  \citet{MartinBouche2009} found that in low-ionization ULIRG outflows, the absorption is saturated at all velocities, while the $C_f$ of the gas decreases with increasing outflow velocity.  This implies that the velocity extent of the detected absorption profile is controlled by the decreasing gas $C_f$ and the $\rm S/N$ of the data, rather than decreasing $N$ at high velocities, and that there may be very low $C_f$ clouds moving at well above the escape velocity.  Higher $\rm S/N$ coadds would be needed to constrain the absorption depth at the escape velocity more precisely.  Hotter, and likely more diffuse, gas kinematics are not constrained by our observations or those of W09.

\subsection{Trends in Outflow Absorption with $\rm \Sigma_{SFR}$} 
As discussed in \S\ref{sec.intro}, \citet{Heckman2002} noted that local galaxies which drive winds strong enough to be detectable in absorption universally satisfy the criterion $\rm \Sigma_{SFR} > 0.1~\rm M_{\odot}~yr^{-1}~kpc^{-2}$.  The galaxies in the W09 sample and LBGs at $z \sim 2-3$ \citep{Shapley2003,Pettini2002} also meet this criterion.  Further evidence for a threshold $\rm \Sigma_{SFR}$ is provided by studies of outflow remnants in emission.  For instance, in a study of the relationship between the sizes of radio halos created by energy input from star formation and the sizes of star-forming regions in local spiral galaxies, \citet{Dahlem2006} found that radio halo size is linearly correlated with the size of star-forming regions.  They additionally found that there is a threshold energy input rate per unit surface area above which radio halos form, and that this threshold depends on the mass surface density of the galaxy.  

There are two different physical mechanisms that may be responsible for driving these winds, 
which result in different dependencies of outflow properties on $\rm \Sigma_{SFR}$.  One mechanism is the thermalization of energy from supernova ejecta, which generates a hot wind that entrains cold clouds in an outflow via ram pressure \citep[e.g.,][]{ChevalierClegg1985,Heckman1990}.  
\citet{McKeeOstriker1977} suggested that a threshold spatial density of supernovae must be achieved in order for this mechanism to operate.  Regions with supernovae densities above such a threshold create and maintain a superbubble filled with hot gas with a cooling timescale much longer than the bubble expansion time \citep{Heckman1990}.  Further, at higher supernova rate per unit volume (or at lower ambient gas density), radiative losses in the bubble are reduced, and the efficiency of thermalization of the ejecta is enhanced \citep{Heckman1990,StricklandHeckman2009}.  This spatial supernova rate density is closely tied to the observable quantity, $\rm \Sigma_{SFR}$.  Additionally, as described by \citet{Martin2005} and \citet{Murray2005}, in this energy-driven paradigm, the terminal velocity of cold clouds in the outflow is proportional to $R_0^{-0.5}$, where $R_0$ is the size of the starburst.  

The other mechanism that may be important in driving winds is momentum from radiation pressure or cosmic ray pressure generated by a starburst or AGN \citep{Murray2005,Socrates2008}.  In the case of radiation-driven outflow, the dependence of outflow velocity on starburst radius for an optically thick wind is slightly weaker, $(\ln 1/R_0)^{0.5}$, while the criterion that must be met to drive a wind depends on the masses of the luminous star clusters driving the outflow.  The star cluster mass, like spatial supernova density, is closely related to $\rm \Sigma_{SFR}$ (see N.~Murray \& B.~M{\'e}nard, in preparation, for further discussion).  Thus, while both proposed mechanisms require that a threshold $\rm \Sigma_{SFR}$ be satisfied to take effect, the empirical relationship between outflow velocity and starburst radius may eventually help constrain the relative contributions of these two mechanisms for driving outflows, although these relationships are complicated by additional factors (e.g., viewing angle, outflow morphology and $C_f$, detection technique and sensitivity).

Our sample suggests that outflow absorption strength increases with $\rm \Sigma_{SFR}$. 
As in \S\ref{sec.disc_sfr_m}, although we measure relatively high values of $W_{diff}$ and outflow EW for our lowest-$\rm \Sigma_{SFR}$ coadd (with $\rm \log \Sigma_{SFR} < -1.36$), because this spectrum has low $\rm S/N$, we consider the detection of outflow tentative, and it must be confirmed in future analyses.  In the middle and high-$\rm \Sigma_{SFR}$ subsamples, the difference between outflow absorption measurements is significant only in the values of $W_{diff}$; the outflow EW values are consistent within the errors.  
It is interesting, though at the limit of our data, to test whether the local criterion on $\rm \Sigma_{SFR}$ for the driving of outflows also applies at $z \sim 1$.  
By accounting for a possible systematic underestimate of SFR by 0.21 dex (see \S\ref{sec.sfr}), we have constructed a subsample with a $\rm \Sigma_{SFR}$ that is likely to be under the local threshold in the mean.  
This coadd has $\rm S/N \sim 4.7~pix^{-1}$, and yields an outflow EW $\sim 1$ \AA.  This value is suggestive of outflow, given the results of our analysis in \S\ref{sec.sn}, although the outflow absorption strength is not as large as in the high-$\rm M_*$ and high-$\rm SFR_{UV}$ subsamples.  However, given the \emph{minimum} estimated dispersion in our SFRs of 0.43 dex, this subsample may include some galaxies with \thresh.  

Additionally, we have measured only a flux-averaged $\rm \Sigma_{SFR}$, with galaxy sizes parametrized by the half-light radius in the UV.  Starburst radii in the local Universe are often constrained using half-light radii measured from UV or H$\alpha$ imaging; however, local and distant starbursts have been found to have distinct morphologies which may make measurements of half-light radii less appropriate for this purpose at $z \sim 1$.  For instance, local LIRGs commonly have centrally concentrated star formation \citep{SandersMirabel1996}, while LIRGs at $z \sim 1$ have more widely distributed star-forming knots \citep{Melbourne2008}.  The half-light radius may therefore not be an accurate characterization of starburst radius in the distant Universe, where it reflects the sizes of the distributions of star-forming knots in the objects, rather than the total surface area of the knots themselves.  Although we have not detected a strong dependence of outflow absorption strength on $\rm \Sigma_{SFR}$, we have by no means ruled it out, and future studies with higher $\rm S/N$ spectra should include more detailed analyses of $\rm \Sigma_{SFR}$ in distant galaxies.

\subsection{Redshift Evolution of Outflow}\label{sec.taus}
\subsubsection{Following the Sample Evolution}
We have identified evolutionary trends in outflow absorption strength which are dependent on galaxy $\rm M_*$.   Specifically, we have found that at $\rm \log M_* \gtrsim 10.49$, outflows are of comparable strength at $z = 1$ and 1.4.  At lower $\rm M_*$, however, there is a significant decrease in outflow absorption strength with decreasing redshift along with a decrease in the SFR in the same galaxies.  Stated in another way, galaxies with similar SFRs at $z=1$ and 1.4 have similarly large outflow absorption strengths, while galaxies with $\rm \log SFR_{UV} < 0.942$ ($\rm < 8.7~M_{\odot}~yr^{-1}$) do not appear to drive strong outflows.  While these conclusions apply to galaxies in specific $\rm M_*$ and $\rm SFR_{UV}$ ranges at different redshifts, they do not speak to the evolution of the outflow properties in a given galaxy through time.
We now attempt to associate the TKRS galaxies with their progenitors in the W09 sample at $z \sim 1.4$ by creating so-called ``$\tau$ models" to parametrize the evolution of SFR and $\rm M_*$ in a given galaxy.  We are largely motivated by the success of $\tau$ models in reproducing the distribution of galaxies in SSFR-$\rm M_*$ space out to $z \sim 1$ \citep{Noeske2007a}.  These models are too simplistic to fully describe the evolution of an individual galaxy, which may experience starbursts, mergers, and other disturbances that define its star formation history.  However, $\tau$ models were intended to describe the broad evolution of populations of galaxies, and are appropriate here, as we are interested in the evolution of large galaxy subsamples.  

In the context of this model, the SFR of a galaxy of total baryonic mass $M_b$ at redshift $z$  is given by 
\begin{eqnarray}
	\mathrm{SFR}(M_b, z) = \mathrm{SFR}(z_f) \exp \left [-\frac{T}{\tau} \right ],
\end{eqnarray}
where $z_f$ is the formation redshift of the galaxy, $T = t(z) - t(z_f)$ is the galaxy age, and $\tau$ is the $e$-folding time.    
\citet{Noeske2007a} found that models with 
\begin{eqnarray}
	\tau(M_b) = c_{\alpha} M_{b}^{\alpha}
\end{eqnarray}
and
\begin{eqnarray}
	1 + z_f (M_b) = c_{\beta} M_{b}^{\beta}
\end{eqnarray}
with the constants $10^{20.4} \leq c_{\alpha} \leq 10^{20.7}$, $10^{-2.5} \leq c_{\beta} \leq 10^{-1.7}$, $\alpha = -1$, and $\beta = 0.3$ successfully reproduce the SSFR-$\rm M_*$ ``main sequence" as well as the spread around this sequence for galaxies from the AEGIS survey at $0.2 < z < 1.1$.  The changing dependence of $z_f$ on $M_b$ is necessary to account for the high SSFRs of galaxies at low $\rm M_*$.  Although these models have not been tested out to $z \sim 1.4$, where the AEGIS sample is incomplete even at the highest $\rm M_*$, we assume that they describe the full locus of star-forming galaxies at this redshift.  

Figure~\ref{fig.taus} shows the location in SSFR-$\rm M_*$ parameter space of the W09 $\rm M_*$-divided subsamples (blue diamonds) and the TKRS $\rm M_*$-divided subsamples (black diamonds).  The point size increases with the square root of the outflow EW for each subsample.  The open diamond indicates the location of the lowest-$\rm M_*$ TKRS subsample, which has insufficient $\rm S/N$ to yield an outflow detection (at 1.2 \AA) with high confidence.  The error bars indicate the spread in galaxy properties in each subsample at the $\pm 34$th percentiles.  
The loci of the SSFR-$\rm M_*$ ``main sequence" at observed redshifts $z_{obs} = 0.35, 1.0$ and 1.4 are shown with nearly horizontal dotted lines from bottom to top.  These loci describe the locations of galaxies in the AEGIS sample of \citet{Noeske2007a} at $z < 1.1$.   
The four solid lines show the path of $\tau$ model galaxies of various $M_b$ (with models for $\log M_b = 10.5, 10.75, 11.0$ and 11.25 plotted from left to right).   
The circles along the single $\tau$ model paths mark the locations of model galaxies with different formation redshifts ($z_f$) at observed redshifts $z_{obs} = 0.35$, 1.0 and 1.4.  The Figure legend gives the appropriate $z_f$ for each color; for instance, purple circles are for the latest $z_f = 1.8$ models, and green circles are for the earliest $z_f = 5.0$ models.    

The W09/DEEP2 galaxies lie slightly above the SSFR-$\rm M_*$ main sequence at $z \sim 1.4$.  This is because the DEEP2 $R$-band selection limits the sample to the highest-SFR galaxies at a given $\rm M_*$; the effect is most severe at $z = 1.4$, the redshift limit of the survey.  (As noted previously, the model for the main sequence at $z \sim 1.4$ has not been observationally tested and is an extrapolation from lower redshift models.)  
The TKRS galaxies, on the other hand, lie below the main sequence at $z \sim 1$.  This discrepancy may be due to a number of factors.  \citet{Noeske2007b} used a combination of $L_{IR}$ and emission line luminosities to constrain the total extinction-corrected SFRs in their sample.  We have shown that our UV-based SFRs are systematically lower (by 0.21 dex) than IR-based SFRs for galaxies 
with IR detections; thus there may be a systematic offset between the SFRs in this work and in \citet{Noeske2007b}.  In addition, we may have failed to remove some red, non-star-forming galaxies from the highest-$\rm M_*$ subsample, which would reduce the median SFR in this bin.  While the offset between the main sequence and our sample is as large as $\sim 0.25$ dex in SSFR for the highest-$\rm M_*$ subsample, this discrepancy does not significantly affect our conclusions, discussed below.

\subsubsection{Outflows in the Sample Progenitors}
We first consider the highest-$\rm M_*$ TKRS subsample.  The median galaxy in this sample has $z_f \sim 5$; i.e., it lies near the green model along the $\rm \log M_b = 11.0$ evolutionary track at $z_{obs} = 1$.  In the case that we are underestimating the sample SFR by $\sim 0.2$ dex, the median galaxy would lie near the cyan model with $z_f \sim 3$.  The highest-$\rm M_*$ W09 subsample appears to include the progenitor of this median galaxy (the green or cyan models at $z_{obs} = 1.4$), as well as the highest SSFR galaxies in the subsample (with later $z_f$).  It appears that outflows persist at similar strengths in high $\rm M_*$ galaxies between $z = 1.4$ and $z = 1$.  Even massive galaxies with early $z_f \gtrsim 5$ 
and very low SSFRs are apparently still driving outflows at $z \sim 1$. 

We next turn to the lower-$\rm M_*$ subsamples.  
The high-SSFR galaxies in the middle-$\rm M_*$ TKRS subsample lie close to the $\log M_b = 10.75$ evolutionary track.  There are W09 galaxies that lie along this track (in the lowest- or middle-$\rm M_*$ W09 subsample), which have SSFRs that may be consistent with progenitors of these high-SSFR TKRS galaxies, particularly if we consider that our SSFR values may be systematically low.  That is, a $\tau$ model galaxy with $\log M_b = 10.75$ and $z_f = 2$ (shown in blue) may be included in the middle-$\rm M_*$ W09 subsample when it is observed at $z_{obs} = 1.4$ and passes through the location of the middle-$\rm M_*$ TKRS subsample at $z_{obs} = 1$.  In the case that we are underestimating the SSFR of this subsample, the middle-$\rm M_*$ galaxies at $z \sim 1$ will lie close to the purple model with $z_f = 1.8$; the progenitors of this model pass through the lowest-$\rm M_*$ W09 subsample at $z_{obs} = 1.4$.
The galaxies with the \emph{lower} SSFR values in this middle-$\rm M_*$ TKRS subsample, however, have no counterpart in W09.  We cannot constrain from this analysis what fraction of the galaxies in this subsample have progenitors in the W09 sample.  However, the W09 galaxies were found to ubiquitously host outflows, while the middle-$\rm M_*$ TKRS subsample does not host strong outflows.  It is unclear whether the outflows at $z \sim 1.4$ have been shut off by $z \sim 1$, or if the lack of outflows in lower-SSFR galaxies is diluting the outflow signature.

The galaxies in the lowest-$\rm M_*$ TKRS subsample may have a few progenitors in the low-$\rm M_*$ end of the lowest-$\rm M_*$ W09 subsample; i.e., there are a few W09 galaxies that lie between the $\log M_b = 10.5$ and 10.75 evolutionary tracks.  However, we do not comment on the presence or absence of outflows in these galaxies at $z \sim 1$, as we lack robust outflow measurements for this subsample due to the low $\rm S/N \sim 2.3 ~pix^{-1}$ of the coadd.

In summary, we find that outflows persist in galaxies with $\rm \log M_* \gtrsim 10.5$ between $z \sim 1.4$ and 1, while in our middle-$\rm M_*$ subsample, outflows may have shut down during this time.  We lack the $\rm S/N$ to comment on the evolution of outflows in the lowest-$\rm M_*$ TKRS galaxies.

\subsubsection{The Sample Descendants}
\citet{Sato2009} investigated outflows traced by \ion{Na}{1} absorption in AEGIS galaxies at $0.11 < z < 0.54$.  Keeping in mind the caveat that \ion{Na}{1} likely traces cooler, higher density gas than \ion{Mg}{2}, we may search for the descendants of TKRS galaxies in the \citet{Sato2009} sample and compare outflow properties.  \citet{Sato2009} are able to measure outflow properties only in galaxies with the highest $\rm S/N$ near the \ion{Na}{1} doublet; these are the highest-$\rm S/N$ spectra in the sample of \citet{Noeske2007a,Noeske2007b}.  As a result, \citet{Sato2009} are complete only down to $\log~ \rm M_* \gtrsim 10.75$, and only at the lowest redshifts ($z \lesssim 0.3$).  Among these galaxies, the vast majority have \ion{Na}{1} absorption at the galaxy systemic velocity, or redshifted with respect to systemic velocity.  At $z > 0.3$, the galaxies with sufficient $\rm S/N$ to measure \ion{Na}{1} kinematics have a mix of outflows, inflows, and absorption at systemic velocity.  These galaxies are massive enough to be descendants of the galaxies in the high-$\rm M_*$ W09 and TKRS subsamples, and yet do not appear to be driving \ion{Na}{1} outflows ubiquitously, particularly at the lowest redshifts.  It may be that either these galaxies are driving outflows of slightly warmer gas with lower density than can be probed by \ion{Na}{1}, or that their outflows have been halted, possibly because the ISM has been removed.  These galaxies have lower SFRs than massive galaxies at $z \sim 1$, consistent with the idea that their cool gas supply has been reduced or exhausted.

In galaxies with $\rm \log M_* < 10.75$, many of which may be descendants of the lower-$\rm M_*$ TKRS subsamples, there is again a range in \ion{Na}{1} kinematics.  Most of the galaxies with kinematic information available are not driving outflows; however, many more have gas kinematics which have not been measured.  Thus the outflow properties of the descendants of the lower-$\rm M_*$ TKRS galaxies remain unconstrained.  

Many of the galaxies in the TKRS sample, like those in the W09 sample, have $\rm \log M_* < 10.7$; i.e., they are below the ``quenching mass" derived by \citet{Bundy2006} for $0.4 < z < 0.7$.  This quenching mass describes the mass at which star formation is suppressed, and decreases from $z = 1.4$ to 0.4.  These lower-$\rm M_*$ galaxies are not driving strong winds for the most part, and because they are well below the quenching limit, are likely to remain blue, star-forming galaxies today.  The TKRS galaxies with $\rm M_*$ above this limit will likely evolve to the red sequence by $z=0$.

\section{Conclusions}
We have analyzed the spectra of \fullsampsize galaxies in the TKRS survey \citep{Wirth2004} of the GOODS-N field at $0.7 < z < 1.5$ and identified cool outflowing gas traced by \ion{Mg}{2} absorption in both individual galaxy spectra and in coadded spectra.  This is the first report of cool outflows in this redshift range.  
We find that the most massive, highest-SFR galaxies in the sample have a strong outflow absorption signature, while less massive 
($\rm \log M_* \lesssim 10.5$, $\rm \log SFR_{UV} < 0.94$) star-forming galaxies have no outflows or outflows with $v \lesssim 100 \mkms$ and with $b_D \lesssim 150 \mkms$.  These same low-$\rm M_*$ galaxies have higher specific SFRs (SSFR) than the high-$\rm M_*$ subsample, but lower absolute SFRs, suggesting that outflow absorption strength is more closely correlated with the latter.

We find that the outflowing gas has a column density $\log N(\mathrm{H}) >  19.3$, resulting in a mass outflow rate of order the SFR of our sample.  This column is slightly smaller than the outflow column densities observed at $z \sim 1.4$ in the DEEP2 survey (W09) and in local IR luminous galaxies \citep[e.g.,][]{Rupke2005a}.  
Our comparison of the \ion{Fe}{2} and \ion{Mg}{2} line profiles in coadded spectra of high-SFR galaxies suggests that \ion{Mg}{2} emission is a common feature of galaxies in this sample.  Although it remains unclear whether the \ion{Fe}{2} absorption in this sample traces outflowing gas, by making the assumption that \ion{Fe}{2} is indeed present in the same outflowing clouds that are traced by \ion{Mg}{2} and that the $C_f$ is independent of outflow velocity, we use the lack of \ion{Fe}{2} absorption blueward of $\rm -300~km~s^{-1}$ to place an upper limit on the outflow column density at these velocities of $\log N(\mathrm{H}) \sim 19.3$.

We also investigate, for the first time, trends in outflow absorption strength with SFR surface density ($\rm \Sigma_{SFR}$) at $z \sim 1$.  We find tentative evidence for a weak trend (at $\sim 1\sigma$ significance) of increasing outflow absorption with increasing $\rm \Sigma_{SFR}$ at $\rm \log \Sigma_{SFR} > -1.36$.  We cannot rule out outflows in galaxies with $\rm \Sigma_{SFR}$ below the local threshold for driving winds, \thresheq, and in fact see some evidence for outflows in these objects.  
Future work in this area should focus on developing a set of measurements of the surface area over which star formation occurs, carefully calibrated total SFRs, and outflow velocities in individual galaxies.

Finally, we examine the temporal evolution of outflows.  We find that galaxies with similar SFRs at $z \sim 1.4$ and $z \sim 1$ have similarly strong outflow absorption.  To examine the evolution of outflows in specific galaxies through time,  
we invoke the $\tau$ model star formation histories of \citet{Noeske2007a}, who note that the decline in SFR at a given $\rm M_*$ with time is consistent with being driven by gas exhaustion.  We find that outflows persist in massive ($\rm \log M_*/M_{\odot} \gtrsim 10.5$) galaxies between $z \sim 1.4$ and $z \sim 1$, in spite of a decreasing SFR between these two epochs.  Measurements of \ion{Na}{1} kinematics at $z \sim 0.2$ \citep{Sato2009} suggest that these massive galaxies, whose SFRs will continue to decline over time, will stop driving winds at low redshift.  We speculate that this is because the cool gas supply in the ISM will be exhausted by $z \sim 0.2$.  Many (but not all) of these galaxies will become red and dead ellipticals by this time.  We measure weak or no outflows (with outflow $v \lesssim 100 \mkms$) in galaxies with lower masses at $z \sim 1$.  Based on our comparison of the SSFR and $\rm M_*$ of our sample to those of $\tau$ model galaxies with exponentially declining star formation histories, we find that the progenitors of some of these lower-$\rm M_*$ galaxies are likely contained in the DEEP2 $z \sim 1.4$ comparison sample and are driving strong outflows at $z \sim 1.4$.  
These objects are likely to remain blue cloud galaxies until $z \sim 0$.  It appears that the outflow strength or the frequency of outflows in this population has declined from $z \sim 1.4$ to $z \sim 1$, as has the SFR.  It is undetermined whether they will ever drive strong outflows as they evolve from $z \sim 1$ to today.

\acknowledgements
K.H.R.R. and D.C.K. acknowledge support for this project from NSF grants AST-0808133 and AST-0507483.
B.J.W. has been supported by NASA/Spitzer contract 1255094.
J.X.P. acknowledges funding though an NSF CAREER grant (AST-0548180).

The authors wish to thank  
J. Melbourne for providing measurements of galaxy radii and IR luminosities, K. Bundy for providing measurements of 
$\rm M_*$, and J. Lotz for providing morphological measurements. 
We thank the TKRS team for making their catalogs and spectra publicly available.  
Finally, we thank Jennifer Lotz, Brice M{\'e}nard, Kai Noeske and David Rupke for interesting and helpful discussions during the analysis of these results.  

The authors wish to recognize and acknowledge the very significant
cultural role and reverence that the summit of Mauna Kea has always
had within the indigenous Hawaiian community.  We are most fortunate
to have the opportunity to conduct observations from this mountain.

{\it Facilities:} \facility{Keck:II (DEIMOS)}, \facility{HST (ACS)}.


\tabletypesize{\scriptsize}
\begin{deluxetable}{lccc}
\tablecolumns{4}
\tablecaption{Measurements of UV absorption line velocity offsets in TKRS2158 \label{tab.veloff2158}}
\tablewidth{0pt}
\tablehead{\colhead{Transition} & \colhead{$\Delta v$}\\ 
	&  \colhead{$\rm km~s^{-1}$}}
\startdata
\ion{Fe}{2} 2344 &  $  -262  \pm     29 $\\
\ion{Fe}{2} 2374 &  $ -281   \pm    42 $  \\
\ion{Fe}{2} 2382 &  $  -415  \pm    36 $  \\
\ion{Fe}{2} 2586  &  $  -287   \pm    40 $\\
\ion{Fe}{2} 2600 & $  -229   \pm    62 $  \\
\ion{Mg}{2} 2796 &   $   -314   \pm    19 $\\
\ion{Mg}{2} 2803 &  $ -257    \pm   30 $\\
\enddata
\tablecomments{Systemic velocities are determined from the velocity of the [\ion{O}{2}] emission line doublet.}
\end{deluxetable}

\begin{deluxetable}{lcccccc}
\tabletypesize{\footnotesize}
\tablecolumns{7}
\tablecaption{EW Measurements of Coadded Spectra\label{tab.ew}}
\tablewidth{0pt}
\tablehead{\colhead{Subsample} & \colhead{$\rm S/N$} & \colhead{$W_{diff}$} & \colhead{Outflow EW} & \colhead{EW(2796)\tablenotemark{a}} & \colhead{EW(2803)\tablenotemark{a}} & \colhead{Doublet Ratio}\\
            &  \colhead{$\rm pix^{-1}$} &  \colhead{ \AA}    &  \colhead{  \AA}   &   \colhead{  \AA}     &   \colhead{ \AA}   &      }
                                     All &  9.72 & $ 0.38 \pm  0.11$ & $ 0.51 \pm  0.14$ & $ 0.51 \pm  0.05$ & $ 0.47 \pm  0.05$ & $ 1.09 \pm  0.16$ \\ 
             $\rm \log SFR_{UV} < 0.555$ &  2.32 & $< 0.94$ & $< 1.39$ & $ < 0.44$ & $  < 0.40$ & \nodata \\ 
     $\rm 0.555 < \log SFR_{UV} < 0.942$ &  6.49 & $ 0.30 \pm  0.17$ & $ 0.33 \pm  0.21$ & $ 0.52 \pm  0.08$ & $ 0.40 \pm  0.07$ & $ 1.29 \pm  0.30$ \\ 
             $\rm \log SFR_{UV} > 0.942$ &  8.26 & $ 1.10 \pm  0.13$ & $ 1.70 \pm  0.17$ & $ 0.78 \pm  0.06$ & $ 0.65 \pm  0.06$ & $ 1.19 \pm  0.14$ \\ 
          ``DEEP2-like" High $\rm SFR_{UV}$ &  6.87 & $ 1.46 \pm  0.16$ & $ 2.48 \pm  0.18$ & $ 0.81 \pm  0.07$ & $ 0.80 \pm  0.07$ & $ 1.01 \pm  0.12$ \\ 
                   $\rm \log M_* < 9.86$ &  3.15 & $ 0.42 \pm  0.35$ & $ 1.20 \pm  0.39$ & $ 0.30 \pm  0.15$ & $ 0.11 \pm  0.15$ & $ < 11.35$ \\ 
           $\rm 9.86 < \log M_* < 10.49$ &  6.07 & $ 0.20 \pm  0.18$ & $ 0.25 \pm  0.22$ & $ 0.44 \pm  0.08$ & $ 0.48 \pm  0.08$ & $ 0.91 \pm  0.22$ \\ 
                  $\rm \log M_* > 10.49$ &  8.04 & $ 1.05 \pm  0.14$ & $ 1.71 \pm  0.18$ & $ 0.81 \pm  0.06$ & $ 0.63 \pm  0.06$ & $ 1.28 \pm  0.16$ \\ 
                              Early Type &  1.90 & $< 1.17$ & $ 1.14 \pm  0.66$ & $ <  0.54$ & $ 0.58 \pm  0.24$ &  \nodata \\ 
                        Merger Candidate &  4.60 & $ 0.47 \pm  0.24$ & $ 0.41 \pm  0.30$ & $ 0.77 \pm  0.11$ & $ 0.69 \pm  0.11$ & $ 1.10 \pm  0.23$ \\ 
                               Late Type &  6.75 & $ 0.33 \pm  0.16$ & $ 0.77 \pm  0.20$ & $ 0.49 \pm  0.07$ & $ 0.56 \pm  0.07$ & $ 0.88 \pm  0.17$ \\ 
         $\rm \log \Sigma_{SFR} < -1.36$ &  3.44 & $ 0.77 \pm  0.31$ & $ 1.15 \pm  0.38$ & $ 0.42 \pm  0.15$ & $ 0.34 \pm  0.13$ & $ 1.21 \pm  0.63$ \\ 
 $\rm -1.36 < \log \Sigma_{SFR} < -0.66$ &  6.83 & $ 0.32 \pm  0.16$ & $ 0.58 \pm  0.20$ & $ 0.56 \pm  0.07$ & $ 0.41 \pm  0.07$ & $ 1.37 \pm  0.29$ \\ 
         $\rm \log \Sigma_{SFR} > -0.66$ &  6.80 & $ 0.84 \pm  0.16$ & $ 0.92 \pm  0.19$ & $ 0.61 \pm  0.07$ & $ 0.63 \pm  0.07$ & $ 0.97 \pm  0.16$ \\ 
  $\rm \log \Sigma_{SFR,Corrected} < -1$ &  4.65 & $ 0.64 \pm  0.24$ & $ 0.90 \pm  0.29$ & $ 0.44 \pm  0.11$ & $ 0.27 \pm  0.10$ & $ 1.64 \pm  0.73$ \\ 
\enddata
\tablenotetext{a}{Measured in velocity interval $\rm -384~km~s^{-1} < v < -200~km~s^{-1}$ relative to the systemic velocity of each transition.}
\tablecomments{Upper limits are given at the $2\sigma$ level.  No doublet ratio is reported if either of the EW measurements is an upper limit.}
\end{deluxetable}

\end{document}